\documentclass[traditabstract]{aa}    
\usepackage{graphicx}
\usepackage[dvips,usenames]{color}
\usepackage{amsmath}
\usepackage{txfonts}
\usepackage{natbib}

\bibpunct{(}{)}{;}{a}{}{,} % to follow the A&A style

\newcommand{\realfigure}[3]{
              \begin{figure}
              \resizebox{\hsize}{!}{\includegraphics{#1}}
              \caption{#2}\label{#3}
              \end{figure}
              }

\newcommand{\mytab}[5]{
\begin{table}
\caption{#4}\label{#5}
\centering
\begin{tabular}{#1}
\hline \hline 
#2
\hline
#3
\hline
\end{tabular}
\end{table}
}

\newcommand{\mytabnote}[7]{
\begin{table}
\caption{#4}\label{#5}
\centering
\begin{#6}
\begin{tabular}{#1}
\hline \hline 
#2
\hline
#3
\hline
\medskip
\end{tabular}
\end{#6}
\begin{flushleft}{#7}\end{flushleft}
\end{table}
}

\newcommand{\nirpapercitep}{\citep{held+2008leo1}}

\newcommand{\deltasigmaW}{\Delta_{\Sigma W}}
\newcommand{\mhdisptrue}{0.08}
\newcommand{\fehdisptrue}{0.14}
\newcommand{\metgradkpc}{-0.27}
\newcommand{\vsys}{271} %= 271.14
\newcommand{\sigmavsys}{13.7} %= 13.73
\newcommand{\radec}{$10^h 08^m 28\fs1$, $+12\degr 18\arcmin 23\arcsec$
  (J2000)}
\newcommand{\numexp}{{100}}

\newcommand{\caii}{\ion{Ca}{ii}}

\newcommand{\noun}[1]{\textit{#1}}

\newcommand{\feh}{{\rm [Fe/H]}}
\newcommand{\mh}{{\rm [M/H]}}
\newcommand{\afe}{{\rm [}\alpha\rm{/Fe]}}
\newcommand{\vhb}{V_\text{HB}}
\newcommand{\kms}{km s$^{-1}$}

\newcommand{\marco}[1]{{{#1}}}

\newcommand{\referee}[1]{#1}         %%  do-nothing

\def\nodata{\ldots\null}

\newcommand{\abbrev}[1]{{#1}}      %%  plain, final

\newcommand{\leoi}{Leo\,I}
\newcommand{\leoii}{Leo\,II}

\begin{document}
\title{New constraints on the chemical evolution of the dwarf
     spheroidal galaxy \leoi\ from VLT spectroscopy
  \thanks{Based on data collected at the European Southern
    Observatory, Paranal, Chile, Proposals No. 69.D-0455 and 71.D-0219}}
%\subtitle{}

   \author{M.~Gullieuszik\inst{1}
          \and
          E.~V.~Held\inst{1}
          \and
          I.~Saviane\inst{2}
          \and
          L.~Rizzi\inst{3}
          }

   \offprints{M. Gullieuszik}

   \institute{
Osservatorio Astronomico di
Padova, INAF, vicolo dell'Osservatorio 5, I-35122 Padova, Italy\\
\email{marco.gullieuszik@oapd.inaf.it,enrico.held@oapd.inaf.it}
        \and
European Southern Observatory, Casilla 19001,
Santiago 19, Chile\\
\email{isaviane@eso.org}
        \and
Joint Astronomy Centre, 660 N. A'ohoku Place, 
University Park, Hilo, HI 96720, USA\\
\email{l.rizzi@jach.hawaii.edu}       
}

  \date{Received \dots; accepted \dots}

 \abstract{
%-aim 
   We present the spectroscopy of red giant stars in the dwarf
   spheroidal galaxy \leoi, aimed at further constraining its chemical
   enrichment history.
%-methods
   Intermediate-resolution spectroscopy in the \caii\ triplet spectral
   region was obtained for 54 stars in \leoi\ using FORS2 at the
   ESO Very Large Telescope.  The equivalent widths of \caii\ triplet
   lines were used to derive the metallicities of the target
   stars on the [Fe/H] scale of Carretta \& Gratton, as well as on a
   scale tied to the global metal abundance, [M/H].
%-results
   The metallicity distribution function for red giant branch (RGB)
   stars in \leoi\ is confirmed to be very narrow, 
   \referee{
     with mean value 
     \mh$\simeq -1.2$ and dispersion $\sigma_\mh \simeq 0.2$.
     By evaluating all contributions to the measurement error, 
 }
   we provide a
   constraint to the intrinsic metallicity dispersion, $\sigma_{\rm
     [M/H],0} = \mhdisptrue$.  We find a few metal-poor stars (whose
   metallicity values depend on the adopted extrapolation of the
   existing calibrations), but in no case are stars more metal-poor
   than [Fe/H]~$ = -2.6$.
   Our measurements provide a hint of a shallow metallicity
   gradient of $\metgradkpc$ dex Kpc$^{-1}$ among \leoi\ red giants.
   The gradient disappears if our data are combined with previous
   spectroscopic datasets in the literature, so that any firm
   conclusions about its presence must await new data, particularly in
   the outer regions.
   By combining the metallicities of the target stars with their
   photometric data, we provide age estimates and an age-metallicity
   relation for a subset of red giant stars in \leoi.  Our age
   estimates indicate a rapid initial enrichment, a slowly rising
   metal abundance --consistent with the narrowness of the metallicity
   distribution-- and an increase of $\sim 0.2$ dex in the last few Gyr.
   The estimated ages also suggest a radial age
   gradient in the RGB stellar populations, which agrees with the
   conclusions of a parallel study of asymptotic giant branch stars in
   \leoi\ from near-infrared photometry.  Together, these studies
   provide the first evidence of stellar population gradients in
   \leoi.  }


   \maketitle
%

%---------------------------------------->
\begin{figure*}
\begin{center}
\includegraphics[clip,width=1.0\textwidth]{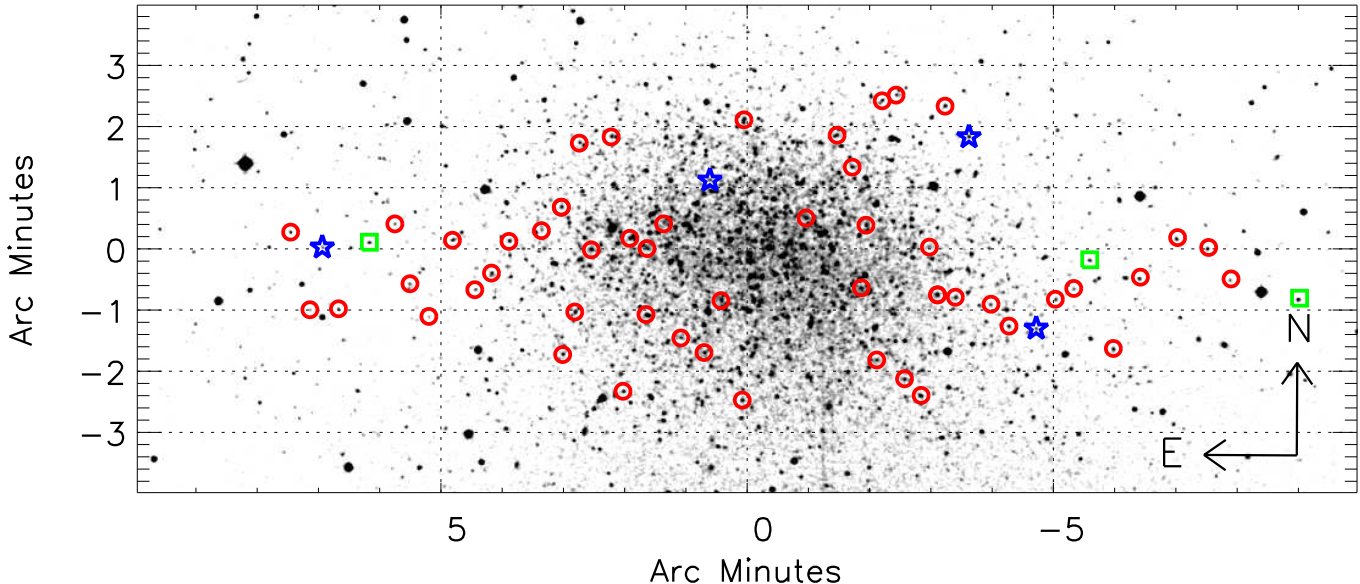}\\
\end{center}
\caption{Digitized Sky Survey image of the \leoi\ field, centred at
\radec.  
{\it Starred symbols} mark the 4 metal-poor member stars, the 
{\it circles} are other member stars, {\it squares} represent interlopers. 
\label{fig:map} }
\end{figure*}
%----------------------------------------|

%%%%%%%%%%%%%%%%%%%%%%%%%%%%%%%%%%%%%%%%%%%%%%%%%%%%%%%%%%%%%%%%%%
\section{Introduction}

%- \subsection{First approach: a photometric MDF}

A first-order estimate of the distribution of stellar metallicities in
resolved galaxies can be obtained from photometry of red giant branch
(\abbrev{RGB}) stars. However, young  metal-rich stars have the same colours as
older metal-poor stars, a phenomenon known as ``age-metallicity
degeneracy''. Although improved photometric metallicity estimates can
be obtained by combining optical and near-infrared photometry, with
reduced degeneracy effects \citep{gull+2007,gull+leo2IR}, the problem
is never entirely overcome, and age and metallicity inextricably
contribute to the observed colour.

%-- \subsection{The importance of spectroscopy}

More direct measurements of the stellar metallicity distribution
function (\abbrev{MDF}) are 
obtained from spectroscopy.  The most accurate determinations come 
from high-resolution abundance analysis which provides 
information on the relative abundances of chemical elements, and
therefore the strongest constraints to the galaxy evolution models
\citep[e.g.][]{tols+2003,lanfmatt2007}.
High-resolution spectroscopy, however, is limited to the brightest
stars in nearby dwarf galaxies, and becomes unfeasible for more
distant galaxies even within the Local Group  and using
10m-class telescopes.
%
%--- low res. spectroscopy:
A viable alternative for obtaining metallicities for a large number of
stars (as needed to derive statistically significant stellar
\abbrev{MDFs}) is using low- or intermediate-resolution spectroscopy.
The infrared Ca triplet (\abbrev{CaT}) method, originally devised to
measure metallicities of stars in Galactic globular clusters
\citep{armazinn1988,armadaco1991,rhs+1997}, 
has now been applied quite extensively
to \abbrev{RGB} stars in dwarf galaxies.  Using this method, values of
\feh\ for dwarf galaxies have been obtained for Fornax \citep{tols+2001,
pont+2004, batt+2006}, Carina \citep{koch+2006}, Sculptor
\citep{tols+2001}, \leoii\  \citep{bosl+2007, koch+2007leo2}, NGC 6822
\citep{tols+2001} and the LMC \citep{cole+2005}.

%- \subsection{Motivations: why Leo\,I ?}

In this paper we present new CaT spectroscopy for the 
dwarf spheroidal (\abbrev{dSph}) galaxy \leoi.
\leoi\ is known to have formed the bulk of its stars at an
intermediate epoch \citep[e.g.,][]{gall+1999}. Along with \leoii,
\leoi\ is one the most distant \abbrev{dSph} satellites of the Milky
Way, for which the influence of tidal interaction with our Galaxy on
evolution must have been more limited than for nearby galaxies such as
Carina or Fornax.  As such, deriving its chemical enrichment history
is of foremost importance for our knowledge of the evolution of
\abbrev{dSph} galaxies.
%
%- \subsection{Previous work: photometric metallicities}
%
\citet{gall+1999} derived the metallicity and reconstructed the
star-formation history (SFH) of \leoi\ from HST/WFPC2 data -- they found a metallicity
ranging from \feh \ $ =-1.4$ to \feh\ $=-2.3$.  Using the same data
set, \citet{dolp2002} found higher metallicity, ranging from \feh\
$=-0.8$ to \feh \ $=-1.2$.   From the colour of \abbrev{RGB} stars,
\citet{held+2000} derived a mean metallicity \feh \ $\sim -1.6$
on the \citet{zinnwest1984} scale.

The metallicity distribution of RGB stars in \leoi\ has recently been
investigated by \citet{bosl+2007} and \cite{koch+2007leo1}.
\citet{bosl+2007} used the CaT method to analyse Keck-LRIS spectra
of 102 RGB stars, and found ${\rm [Fe/H]}=-1.34$ on the \feh\
scale of \citet{carrgrat1997} (hereafter, \abbrev{CG97}).  The authors
also proposed a new calibration based on Ca abundance, yielding a
mean metallicity $\text{[Ca/H]} = -1.34$ ($\sigma = 0.21$).
Also using measurements of the \caii\  triplet lines,
\cite{koch+2007leo1} found a mean metallicity \feh\ $=-1.31$ on the
CG97 scale for 58 red giants.
In both studies, the \abbrev{MDF} is well described by a Gaussian
function with a 1$\sigma$ width of 0.25 dex, and a full range in \feh\
of approximately 1 dex.

% population-gradient

%
\leoi\ is an interesting target also because, to date, it is one of the
few Local Group dwarf galaxies showing scarce evidence of a
population gradient.  \citet{held+2000} found that the old horizontal
branch stars of Leo I are radially distributed as the intermediate-age
helium-burning stars; \citet{koch+2007leo1} found no significant
metallicity radial gradient.
Different conclusions were found in our companion paper, based on
near-infrared photometry \citep{held+2008leo1}, showing that
intermediate-age asymptotic giant branch (AGB) stars are more
concentrated in the central region than old RGB stars.

A new, independent data set of metallicities for \leoi\ stars, also
based on the \abbrev{CaT} method, was obtained by us at the ESO VLT
using high signal-to-noise FORS2 spectra.  The new sample has
negligible overlap with the previous data sets, thus effectively
increasing the number of \leoi\ stars with direct metallicity
measurements.  Using the new data, this paper provides an
independent determination of the \abbrev{MDF} of \leoi\ and further
constraints on its evolution, based on an analysis of metallicity and
age gradients and the age-metallicity relation.

%%%%%%%%%%%%%%%%%%%%%%%%%%%%%%%%%%%%%%%%%%%%%%%%%%%%%%%%%%%%%%%%%%
\section{Observations and reduction}
\label{s:obsredu}

%-----------------------------
\subsection{Target selection}

Our targets were selected from the
colour-magnitude diagram (\abbrev{CMD}) of \leoi. For the central region
of the galaxy, we relied upon  $B$,$V$-band photometry from
\citet{held+2000}, obtained with the EMMI instrument at the NTT
telescope at the ESO La Silla Observatory. For the stars in the outer
regions, we used the $BV$ photometry originally obtained for a study of
RR~Lyrae variable stars in \leoi\ \citep{held+2001}, based on
observations carried out with the Wide Field Imager at the 2.2m ESO-MPI
telescope.

The spectroscopic targets were selected among the
brightest RGB stars of \leoi, down to 1 mag below the RGB tip.  We
avoided any colour constraints that might bias the
age/metallicity distribution. After mask design (for which we were guided only by
geometric constraints) we were left with 61 targets in 4 masks.

The identifiers, coordinates, and $BV$ photometry of the stars in our
final sample (excluding a few targets with too low S/N ratio to allow
any measurements) are listed in Table~\ref{t:data2}, and the targets
are shown in Fig.~\ref{fig:map}.  The location of the target stars in
the CMD of \leoi\ is shown in Fig.~\ref{fig:cmd}.

%------------------------------------>
\realfigure{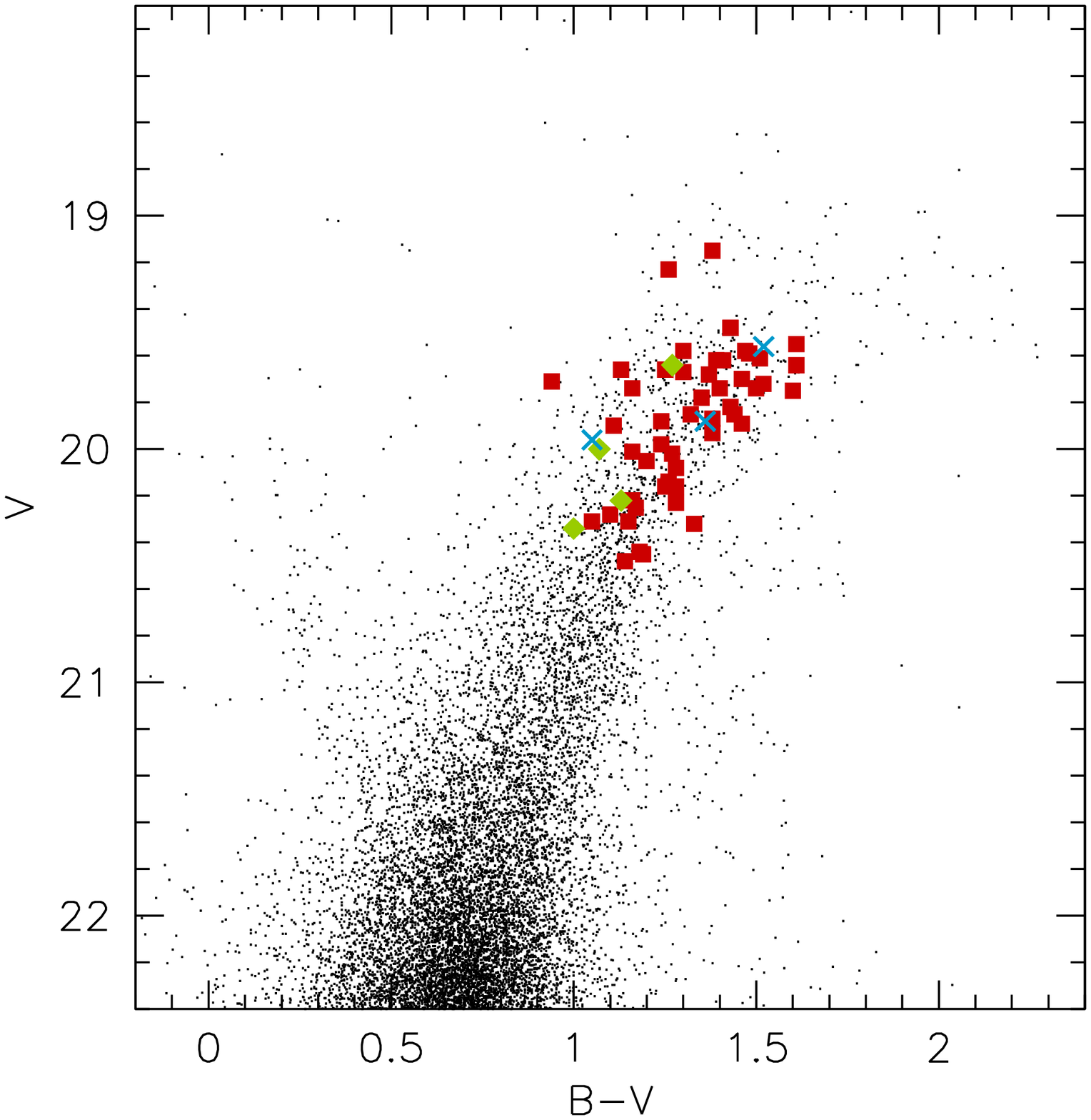}{Target stars in the colour-magnitude diagram of
\leoi. The {\it filled squares} represent red giant stars in \leoi, 
the {\it diamonds} are the 4 metal-poor red giant members, 
while {\it crosses} mark non-members. }{fig:cmd}
%------------------------------------|

%-----------------------------
\subsection{Observations}

\addtocounter{table}{1}

\mytab{l r@{ } l@{ } l l}
{
Field&
\multicolumn{3}{c}{Night}&
$t_\text{exp}$ (sec)\\
}{
NGC~4590 & 6       &May &2002 &$15+60$\\
NGC~5927 & 6       &May &2002 &$60+300$\\
NGC~6171 & 6       &May &2002 &$10+60$\\
NCG~5904 & 3       &May &2003 &$60+300$\\
NGC~6397 & 5       &May &2003 &$60+300$\\
NGC~6528 & 5       &May &2003 &$60+300$\\
NGC~4372 & 24       &May &2003 &$60+300$\\
NGC~6752 & 24       &May &2003 &$60+300$\\[2pt]
\leoi~field~U     & 4       &May &2003 &$2\times2870$ \\
\leoi~field~R     & 20,21 &Dec &2003 &$2\times2870$ \\
\leoi~field~D     & 22       & May &2003 &$2\times2870$ \\
\leoi~field~L     & 16,25 & Jun &2003  &$2\times2870$  \\
}
{Log of the observations.}{tab:log}

The observations were carried out in service mode in two runs
between May 2002 and December 2003 using FORS2,
the multi-mode optical instrument mounted on the 
Cassegrain focus of the Yepun (VLT-UT4) 8.2m telescope at the ESO
Paranal Observatory.  We used FORS2 in MXU mode with the 1028z+29 grism
and the OG590+32 order-blocking filter. With this setup, the spectral
coverage is approximately 7700 \AA\ to 9500 \AA, with a dispersion 
0.85 \AA\ pixel$^{-1}$.  The selected targets were observed 
with 4 masks using $0\farcs80$ slits.  For each \leoi\ mask, two
spectra were taken.  The observing log and exposure times are given in
Table~\ref{tab:log}.
We observed also RGB stars in 8 Galactic globular clusters
(\abbrev{GCs}) in a wide range of metallicity with the
same instrumental setup in order to calibrate our measurements onto a
known metallicity scale. Two different exposure times were used (long
and short exposure) to prevent saturation of the brightest
\abbrev{RGB} stars.
The photometry of the \abbrev{GC} stars was taken from the data
compilation of \citet{rhs+1997}.

%------------------------------------------------>
\realfigure{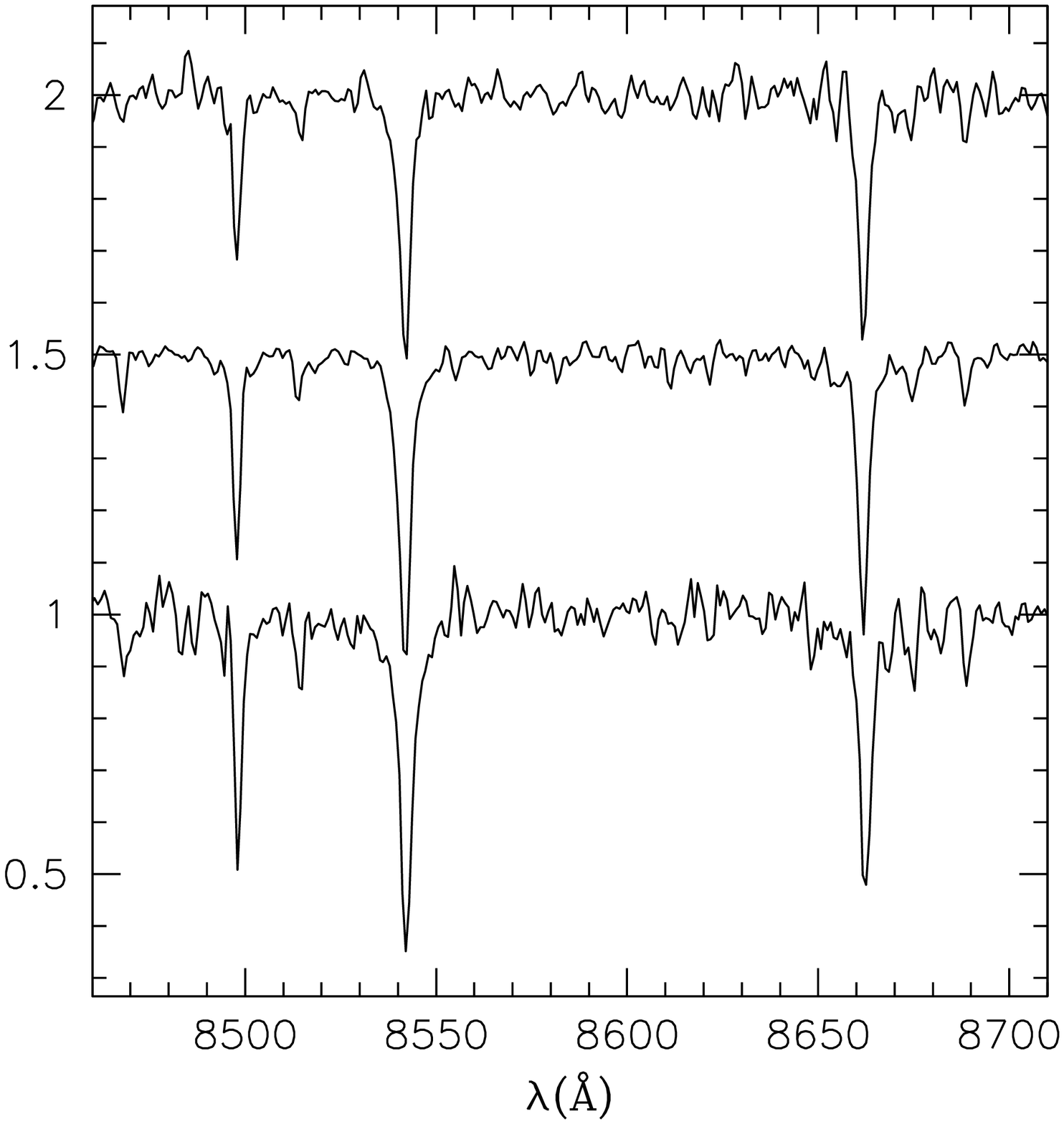}{Examples of normalised,
background-subtracted spectra of RGB stars in \leoi\ (stars \#17, \#19,
\#54, from top to bottom).  The spectra are shown on a rest-frame wavelength
scale and vertically shifted for clarity.}{fig:spec}
%------------------------------------------------|

\subsection{Data Reduction}

The basic reduction of multi-object spectra was performed using standard
procedures in IRAF
\footnote{
  IRAF is distributed by the National Optical Astronomy Observatory,
  which is operated by the Association of Universities for Research in
  Astronomy (AURA) under cooperative agreement with the National
  Science Foundation.}.
Bias and flat-field corrections were applied to all images using the
\noun{ccdproc} task. Due to the long exposure times, the scientific
frames for \leoi\ contain a large number of cosmic ray hits. These were
effectively cleaned using the IRAF program \noun{lacos} \citep{vand2001}
on the bias-subtracted, flat-fielded images.  The multi-object spectra
were extracted with the \noun{apall} task and wavelength calibrated
using HeNeAr lamp exposures taken at the end of each night.

The two spectra taken for each \leoi\ target were combined to increase
the S/N ratio. However, we also retained the two individual spectra to
estimate the uncertainties in the wavelength calibration and  line
strengths.  Finally, the continuum was normalised in the region between
8400 \AA\ and 8800 \AA, by excluding in the process the \caii\ and other
relatively strong absorption lines. Typical sky-subtracted spectra
of three \leoi\ stars with different metallicities are shown in
Fig.~\ref{fig:spec}.

\referee{ The average signal-to-noise ratio per pixel was calculated
  from the rms of the combined spectra in two wavelength windows free
  from strong spectral features, 8580--8620 \AA\ and 8710--8750 \AA.
  The values were checked against those measured on the raw spectra,
  and found to be consistent within a few percent. The S/N ratio, listed
  in Table~\ref{t:data2}, is $\gtrsim 20$ for all stars in our sample,
  with a mean value of $\sim50$.  }

%%%%%%%%%%%%%%%%%%%%%%%%%%%%%%%%%%%%%%%%%%%%%%%%%%%%%%%%%%%%%%%%%%
\section{Radial velocities and membership}
\label{s:leospecvrad}

%----------------------------------------------->
\realfigure{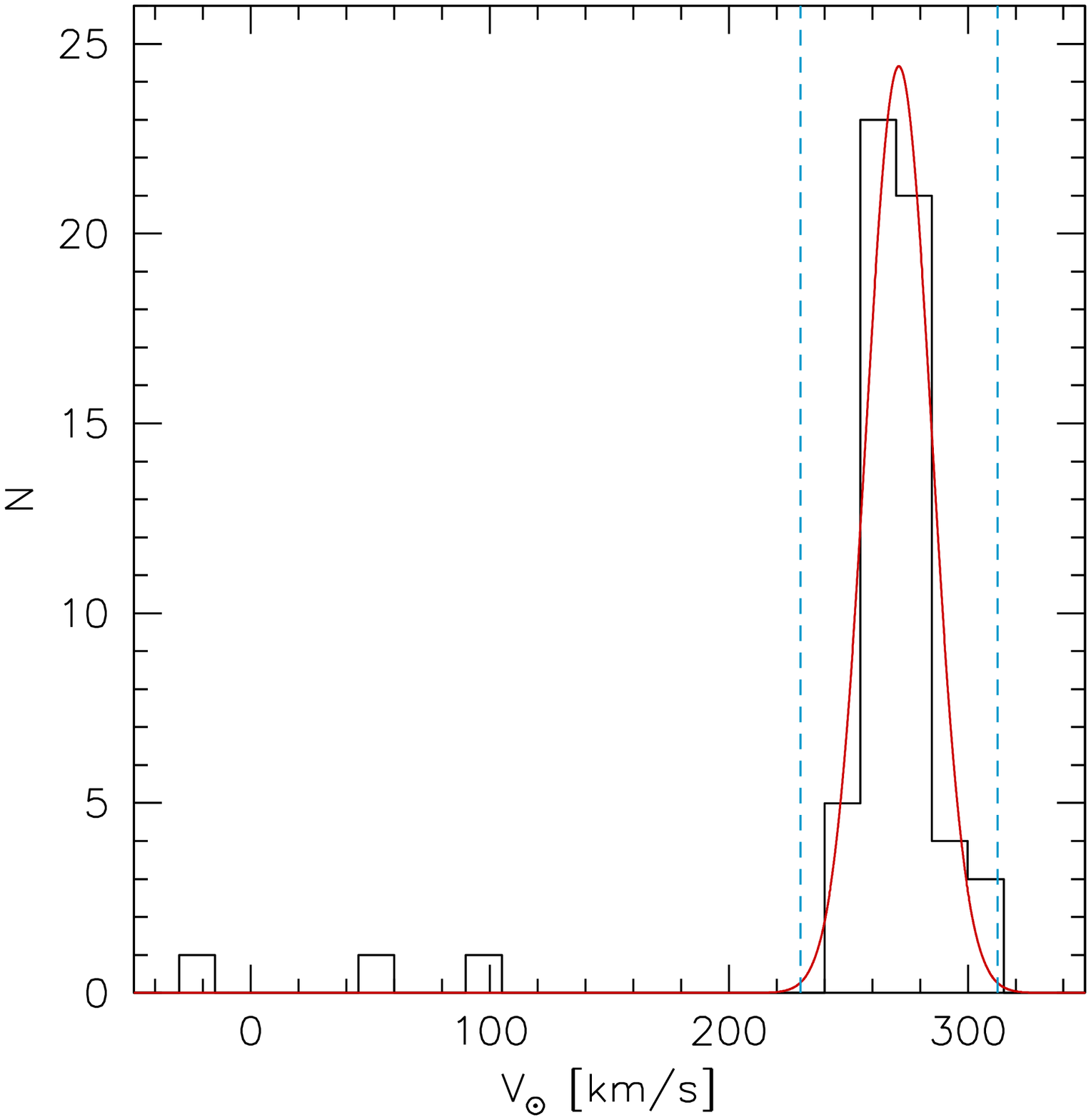}{
  Heliocentric radial velocities of stars in the \leoi\ field.  The
  distribution is fitted with a Gaussian centred at \vsys\ \kms and
  with dispersion \sigmavsys\ \kms. The vertical dashed lines
  represents the $3\sigma$ limits used to select members of \leoi.
}{f:vrad}
%-----------------------------------------------|

Radial velocities were measured for target stars to establish their
membership.  Only the two reddest lines were used to this purpose, since
the bluest line of the Ca triplet is weak and, for the systemic velocity
of \leoi, overlapping with a strong sky line.
The line wavelengths were obtained from the central values $\lambda_m$
of the fitted profile (see Eq.~\ref{e:linefit} below), and compared with
the laboratory air wavelengths 8542.09 \AA\ and 8662.14 \AA.  
A radial velocity was measured for each individual spectrum of
each star by combining the measurements of the two lines
$\lambda_{8542}$ and $\lambda_{8662}$.  
Then, the radial velocity was calculated as the mean of the
two values independently measured from the single spectra.  The
results, corrected to heliocentric velocities using the 
\noun{rvcorr} task, are given in Table~\ref{t:data2}.
The distribution of radial velocities, shown in Fig.~\ref{f:vrad}, is
well fitted by a Gaussian function centred at \vsys\ \kms\ with a dispersion 
\sigmavsys\ \kms.  
All but three stars have heliocentric radial velocities within
$3\sigma$ of the peak, and therefore are considered members of \leoi\ 
(members have ID $< 100$ in Table~\ref{t:data2}). 

Radial velocity errors (given in Col.~7 of Table~\ref{t:data2}) were
estimated by calculating the differences between radial velocities
measured on the two individual spectra.
The distribution of the differences is approximately Gaussian with a
mean $-6.4$ \kms \ and dispersion 10.4 \kms.  The mean velocity
difference provides a good estimate of the mask-to-mask systematic
errors, which include wavelength calibration errors and the centring
errors of the targets on the slitlets, while the standard deviation
represents a combination of random errors.  We assume
$\sigma_v=\sigma_{v_1-v_2}/\sqrt{2}=7.4$ \kms\ as a good approximation
to the standard error of the radial velocity measured on the combined
spectrum.
Therefore our systemic radial velocity of \leoi\  is
\vsys~$\pm \, 6.4$ (systematic) $\pm \, 7.4$ (random) \kms.
\marco{
Our error estimates are consistent with the 0.85 \AA\ pixel$^{-1}$
spectral resolution provided by our instrumental setup, which
corresponds to a resolution in radial velocity of $~\sim 30$ \kms  pixel$^{-1}$
in the CaT wavelength range.
}
\citet{koch+2007leo1} measured 284.2 \kms\ with a velocity dispersion
of 9.9 \kms, while $282.6 \pm 9.8$ \kms\ was the value measured by
\citet{bosl+2007}.  Most recently, \citet{mate+2008} obtained a mean
heliocentric velocity $282.9 \pm 0.5$ \kms\ and a dispersion $9.2 \pm
0.4$ \kms\ from echelle spectroscopy of 328 \leoi\ members.
The systematic difference of $\sim 10$ \kms\ between our
mean velocity estimate and previous results is 
consistent with pointing errors.
In the CaT wavelength range, a velocity of 10 \kms\ corresponds to 
a shift of $\sim 1/3$ pixel on the detector (see above),
or $0\farcs08$ on the sky.

%%%%%%%%%%%%%%%%%%%%%%%%%%%%%%%%%%%%%%%%%%%%%%%%%%%%%%%%%%%%%%%%%%
\section{Equivalent widths and metallicity}
\label{s:leospecmetallicity}

%-----------------------------------
\subsection{Equivalent width measurements}

%-------------------------------------------------------> 
\realfigure{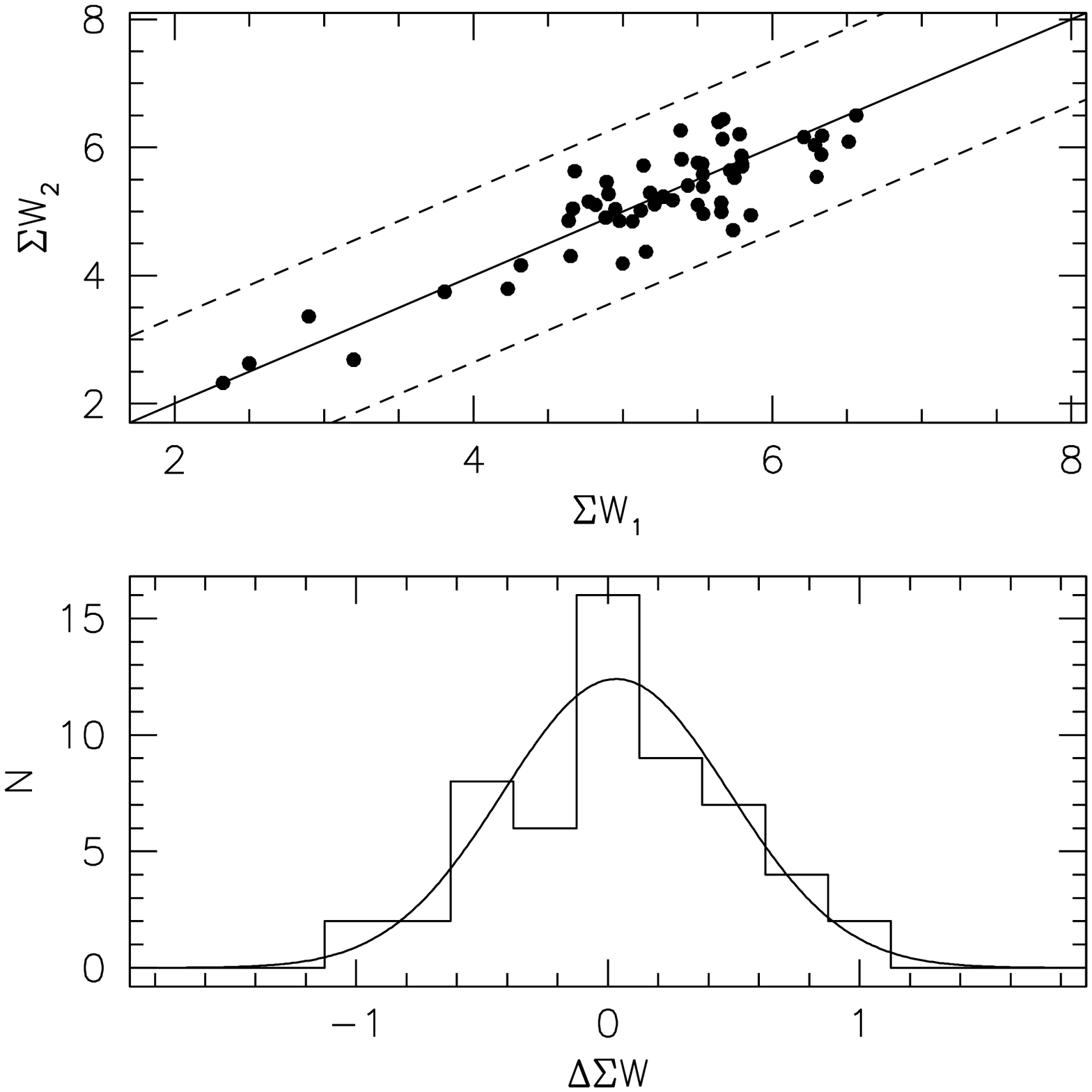}{
  {\textit{Upper panel:}} correlation between two independent EW
  measurements on individual spectra of \leoi\ stars. The {\it dashed
    lines} show the $3\sigma$ interval. {\textit{Lower panel:}}
  histogram of the differences between the two measurements, fitted by
  a Gaussian profile with $\sigma_{\Delta\Sigma W}=0.44$ \AA.
}{fig:err}
%-------------------------------------------------------|

We measured the equivalent widths (\abbrev{EWs}) of \abbrev{CaT}
lines in the spectra of target stars in \leoi\ and the calibrating
GCs as follows.
We first normalised the spectra over a wavelength
interval encompassing, for each line, the side bands defined by
\citet{armadaco1991}. Equivalent widths were then measured for the two
stronger CaT lines in the co-added spectra by fitting a model profile
over the line central bandpasses as defined by the same authors.
Following \citet{cole+2004}, the fitted model is the sum of a
Gaussian and a Lorentzian profiles with a common line centre $\lambda_m$,

\begin{equation}\label{e:linefit}
F(\lambda)=1
-A_G\exp \left[ -\frac{(\lambda-\lambda_m)^2}{2\sigma^2}  \right]\\
-A_L \left[ \frac{(\lambda-\lambda_m)^2}{\Gamma^2}+1\right]^{-1}
\end{equation}
with the best-fit parameters determined using a Levenberg-Marquardt
least-squares algorithm (coded in an {\sc idl} procedure by
C. Markwardt\footnote{
\textrm{http://cow.physics.wisc.edu/$\sim$craigm/idl/idl.html}}). 
The \caii\ line strength was then 
defined as the unweighted sum of the two equivalent widths, 
%
%------------------------------------------>
\begin{equation}\label{e:sigmaw}
\Sigma W = EW_{8542} + EW_{8662}
\end{equation}
%------------------------------------------|
%
The sum equivalent widths are given in Table~\ref{t:data} for the red
giants in \leoi\ and in Table~\ref{tab:misureGCS} for the RGB stars in 
the template GCs.

%%- SIGMA_W UNCERTAINTIES:
%
To estimate the equivalent width measurement errors, we also measured
the CaT line strengths independently on the individual spectra of each
star in \leoi. A comparison of the two $\Sigma W$ measurements is shown
in Fig.~\ref{fig:err}, where they appear to be well correlated.
The histogram of the differences $\deltasigmaW=(\Sigma W_2-\Sigma W_1)$
(shown in Fig.~\ref{fig:err}, lower panel) is well fitted by a Gaussian
function
centred at $\langle \deltasigmaW \rangle=0.03$ and with a standard
deviation $\sigma_{\Delta\Sigma W}=0.44$ \AA. Since the two individual
spectra have comparable $S/N$ ratio, we adopt
$\sigma_{\Sigma W}=\sigma_{\Delta\Sigma W}/\sqrt{2} = 0.31$ 
\AA\ as our error estimate for $\Sigma W$ measured on the combined
spectrum. For comparison, the half-range $\epsilon\,_{\Sigma W}=\left\vert \deltasigmaW / 2
\right\vert$ is listed for each \leoi\ star in Table~\ref{t:data}.

\addtocounter{table}{1} % NB. !!!
%-----------------------------------------------------> 
\mytabnote{%put the column formatting -- r@{\hspace{10pt}}
l l c c c
}{%put the tab header
\multicolumn{1}{l}{Cluster}&
ID&
$V$&
$B-V$&
$\Sigma W $(\AA)\\
}{%put the tab body
M~5    & II-45  & 14.75 &  0.82 &  4.62\\ 
       & II-50  & 13.92 &  0.96 &  5.12\\ 
       & II-51  & 14.05 &  0.96 &  4.90\\ 
       & II-80  & 14.31 &  0.91 &  4.89\\ 
       & II-74  & 13.82 &  1.01 &  5.05\\ 
       & I-2    & 13.87 &  1.02 &  5.24\\ 
       & I-50   & 13.91 &  0.97 &  5.06\\ 
       & I-61   & 13.37 &  1.17 &  5.40\\ 
       & I-68   & 12.37 &  1.52 &  6.25\\ 
       & I-71   & 13.01 &  1.29 &  5.57\\ 
%[.1cm]
NGC~4372  & 20     & 12.88 &  1.57 &  3.78\\ 
       & 14     & 14.29 &  1.29 &  2.88\\ 
       & 13     & 12.72 &  1.73 &  4.13\\ 
       & 10     & 13.82 &  1.29 &  3.00\\ 
       & 95     & 14.48 &  1.30 &  2.72\\ 
       & 91     & 14.45 &  1.33 &  2.81\\ 
       & 89     & 14.49 &  1.29 &  2.83\\ 
       & 141    & 12.93 &  1.65 &  4.01\\ 
       & 74     & 14.17 &  1.40 &  3.09\\ 
       & 76     & 14.18 &  1.30 &  3.16\\ 
       & 77     & 14.19 &  1.28 &  3.07\\ 
%[.1cm]
NGC~6171  & 62     & 13.97 &  1.62 &  5.83\\ 
       & 100    & 14.21 &  1.40 &  5.50\\ 
       & I      & 13.89 &  1.46 &  5.66\\ 
       & F      & 13.39 &  1.70 &  6.21\\ 
%[.1cm]
NGC~6397  & 328    & 12.07 &  0.93 &  3.19\\ 
       & 326    & 12.78 &  0.89 &  2.89\\ 
       & 337    & 12.58 &  0.90 &  2.83\\ 
       & 343    & 11.42 &  1.13 &  3.34\\ 
       & 361    & 11.67 &  1.08 &  3.18\\ 
%[.1cm]
NGC~6528  & R2-8   & 15.79 &  1.89 &  6.80\\ 
       & R1-42  & 16.46 &  1.62 &  6.34\\ 
       & R2-41  & 16.30 &  1.64 &  6.44\\ 
%[.1cm]
NGC~6752  & 4      & 13.70 &  0.85 &  3.86\\ 
       & 8      & 11.96 &  1.05 &  4.99\\ 
       & 28     & 13.17 &  0.89 &  4.18\\ 
       & 29     & 11.79 &  1.17 &  4.98\\ 
       & 30     & 12.15 &  1.12 &  4.83\\ 
%[.1cm]
NGC~5927  & 133    & 14.75 &  1.97 &  6.45\\ 
       & 372    & 14.66 &  2.11 &  6.28\\ 
       & 335    & 14.44 &  1.94 &  6.63\\ 
       & 190    & 14.29 &  2.02 &  6.71\\ 
       & 65     & 14.64 &  1.92 &  6.58\\ 
%[.1cm]
NGC~4590  & 144    & 12.80 &  1.29 &  3.72\\ 
       & 239    & 14.19 &  0.87 &  2.88\\ 
       & II72   & 15.03 &  0.85 &  2.60\\ 
       & 30     & 14.15 &  0.87 &  2.64\\ 
       & 74     & 14.59 &  0.84 &  2.36\\ 
       & 119    & 13.62 &  0.95 &  2.90\\
}
{
%put the caption
Observed stars in template globular clusters.
}{%put the label
  tab:misureGCS}{%size
  normalsize}{ {\it Notes}.  The IDs, magnitudes, and colours of stars
  are those given in the original photometry papers quoted by
  \citet{rhs1997}.
  % The sum equivalent widths of CaT lines are given in Col.~5.
}

%------------------------------------------------------------>
\realfigure{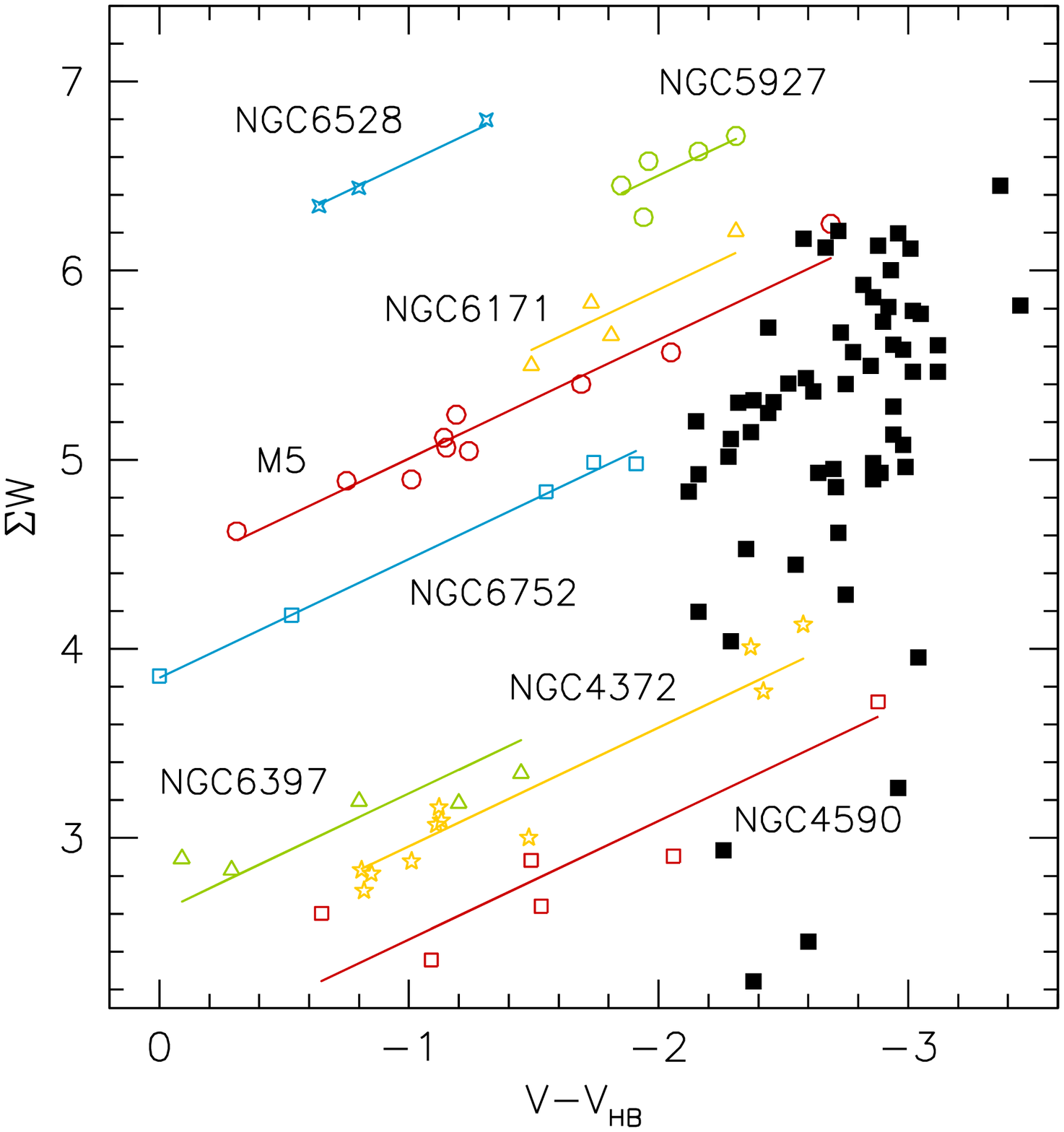}{The sum of the equivalent widths of the two
  most reliable CaT lines plotted as a function of the magnitude
  difference from the HB level.  The calibration globular
  clusters are represented by {\it open symbols} (coded with different colours
  in the electronic edition of the journal).  The metallicity of the
  clusters increases from the bottom to the top (see 
  Table~\ref{tab:paramGCS}).  The straight lines are 
  best fits to the EWs for each cluster, assuming a common best-fit
  slope.  {\it Filled squares} 
  refer to \leoi\ RGB stars.  }{fig:The-sum-of}
%------------------------------------------------------------|

%----------------------
\subsection{Reduced EW}

According to the CaT method \citep{armadaco1991}, 
\referee{the gravity and  $T_{\rm eff}$\  dependence} 
of CaT lines is accounted for by introducing a linear correction to the
line strengths which depends on the star's luminosity, that is 
%---------------------------------->
\begin{equation}\label{e:calib}
W^\prime=\Sigma W+\beta (V-\vhb)
\end{equation}
%----------------------------------|
%
where $\beta$ is a constant and $(V-\vhb)$ is the difference between
the magnitude of the star and the horizontal branch (HB) in the $V$ band.
In globular clusters, this {\it reduced equivalent width} $W^\prime$ was
found to be well correlated with metallicity
\citep{armazinn1988,armadaco1991,rhs+1997}.  This provides the empirical basis
for the validity of the CaT method. Applied to
composite stellar populations, the method is less straightforward and has been
widely discussed in the recent literature.  We will return to this
point later on.

%: sumW vs. V-Vhb 
In Fig.~\ref{fig:The-sum-of}, we plot the sum of equivalent widths 
$\Sigma W$ versus $V-V_{{\rm HB}}$ for all stars with good S/N spectra,
both in \leoi\ and the template \abbrev{GCs}.  The magnitude of the
\abbrev{HB} (of old stars) in \leoi, $V_{{\rm HB}}=22.60$, is from
\citet{held+2001}, while for the template GCs, $\vhb$ was taken from
\citet{rhs1997} (listed in Table~\ref{tab:paramGCS} together with
clusters' metallicities).  For all the globular clusters, our CaT line
strengths define clean, well separated linear sequences generally
consistent with a constant slope and having different,
metallicity-dependent zero points.  
In this diagram, the \leoi\ stars show quite a large dispersion,
although most of them are located between the sequences of NGC\,6397 and
M\,5.  This spread in CaT line strengths is real, being larger than the
typical measurement error $\sigma_{\Sigma W} \approx 0.3$ \AA.
By assuming a common slope and lumping together the data for all
globular clusters, we derived a slope $\beta = 0.627 \pm 0.021$. 
This value can only be compared with previous results that use
the same definition of $\Sigma W$.  This is the case for
\citet{tols+2001}, who found $\beta=0.64 \pm 0.02$, in 
agreement with our result.

%-----------------------------------
\subsection{Metallicity calibration}
\label{s:metcalib}

%------------------------------------------------------------->

\mytab{lcccc}
{
\multicolumn{1}{l}{Cluster}&
\feh$_{\rm ZW}$&
\feh$_{\rm CG}$&
\mh&
$\vhb$\\
}{
%-NGC~6528         &$-0.23$ &$-0.10$ &$-0.03$ &15.70 \\
NGC~6528         &$-0.23$ &$-0.10$ &$-0.03$ &17.10 \\
NGC~5927         &$-0.31$ &$-0.46$ &$-0.37$ &16.60 \\
%-NGC~6171 (M107)  &$-0.99$ &$-0.87$ &$-0.70$ &13.70\\
NGC~6171 (M107)  &$-0.99$ &$-0.87$ &$-0.70$ &15.70\\
NGC~5904 (M~5)   &$-1.40$ &$-1.11$ &$-0.90$ &15.06\\
%-NGC~6752         &$-1.54$ &$-1.42$ &$-1.21$ &17.10\\
NGC~6752         &$-1.54$ &$-1.42$ &$-1.21$ &13.70\\
NGC~6397         &$-1.91$ &$-1.82$ &$-1.65$ &12.87\\
NGC~4372         &$-2.08$ &$-1.94$ &$-1.74$ &15.30\\
NGC~4590 (M68)   &$-2.09$ &$-1.99$ &$-1.81$ &15.68\\
}{
Parameters for the calibration Galactic globular clusters.}
%{Metallicities are from \citet{ferr+1999}, except for NGC\,6528, whose
%values have been updated according to the results of
%\citet{zocc+2004}. The mean $V$ magnitude of the HB is from
%\citet{rhs1997}.}
{tab:paramGCS}

%-------------------------------------------------------------|

%-------------------------------------------------------> 
\mytab{l l @{= } r@{ } r@{ } r@{ }}
{
Type&
\multicolumn{4}{c}{Calibration}\\
}{
%---
quadratic&
 \feh$^{\text ZW}$&$0.088 \,{W^\prime}^2$&$-0.184 \, W^\prime$&$-2.079$\\
&\feh$^{\text CG}$&$0.072 \,{W^\prime}^2$&$-0.076 \, W^\prime$&$-2.122$\\
&\mh             &$0.051 \,{W^\prime}^2$&$+0.056 \, W^\prime$&$-2.125$\\
linear&
 \feh$^{\text ZW}$  &&$0.359\, W^\prime$&$-2.845$\\
&\feh$^{\text CG}$  &&$0.391\, W^\prime$&$-2.806$\\
&\mh               &&$0.395\, W^\prime$&$-2.628$\\
}
{
Metallicity calibrations.
%The relations between the reduced EW of CaT lines in
%Galactic globular clusters stars, $W^\prime$, and the known
%metallicities of the clusters on three different metallicity
%scales. Both quadratic fits in the whole metallicity range and linear
%fits for clusters with $\text{[Fe/H]} < -0.6$ are given.
}{tab:fit}
%-------------------------------------------------------|

%---------------------------------------------------------------->
\realfigure{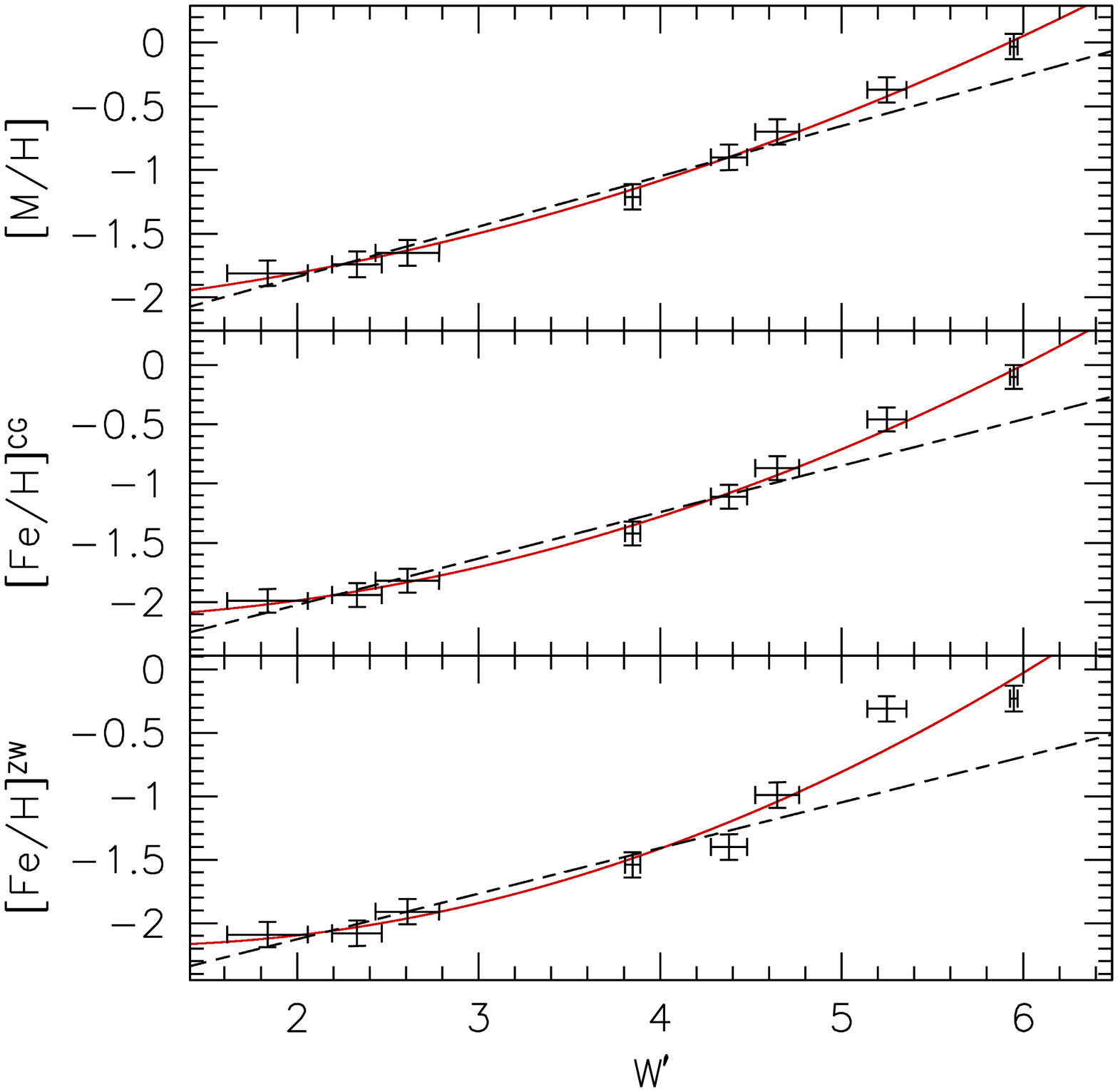}{The metallicity of the reference
  globular clusters against the reduced EW of CaT lines, for three
  adopted metallicity scales.  The {\it solid} lines are quadratic fits,
  while the {\it dashed} straight lines are linear fits obtained for
  clusters with \feh $<-0.6$.}{fig:WvsFe}
%----------------------------------------------------------------|

Using our reduced CaT equivalent widths $W^\prime$ and the published
metallicities for the Galactic globular clusters, we re-determined the
calibration relations between $W^\prime$ and metallicity on 3 different
abundance scales: the $\feh$ scales of \citet{zinnwest1984} and
\citet{carrgrat1997} and the {\it global} metallicity $\mh$, as defined
by \citet{sala+1993}.
The metallicities of the GCs (Table~\ref{tab:paramGCS}) were taken from
\citet{ferr+1999}, except for the metal-rich cluster NGC\,6528, for
which the more recent results of \citet{zocc+2004} were adopted
($\feh=-0.1$, $\afe=0.1$).  The global metallicity of NGC\,6528
($\mh=-0.03$) was calculated using the relation from
\citet{sala+1993}:%
\begin{equation}
\begin{split}
\mh=&\feh+\log (0.683 f_\alpha+0.362)\\
\log f_\alpha=&\afe
\end{split}
\end{equation}

Figure~\ref{fig:WvsFe} shows the $W^\prime$-metallicity relations for
the three scales, along with quadratic fits to the whole dataset and
linear fits to the metal-poor and intermediate globular clusters.
The quadratic relations provide a better fit to the GC metallicities
over the whole metallicity range of template GCs.  The curvature is
driven by the data for two most metal-rich globular clusters,
consistently with a fall in sensitivity of the \caii\ index at high
metallicity.  Previous studies which included metal-rich GCs also
found quadratic relations \citep{armadaco1991, dacoarma1995, carr+2001,
bosl+2007}.
Linear relations have been proposed by other studies \citep[most
recently, ][]{cole+2004,koch+2007leo1,carrera+2007} using metal-rich
open clusters to constrain the metal-rich end of the $W^\prime$ --
[Fe/H] relation.
A full discussion of the behaviour of CaT line strengths against
metallicity is beyond the scope of this paper, and will be presented in a
future paper along with a large dataset of calibrating globular
clusters.
For our data, a linear relation indeed provides a good fit for stars
less metal-rich than the template cluster NGC\,6171 (\mh\ $=-0.70$)
(Fig.~\ref{fig:WvsFe}).
In the case of a metal-poor system such as \leoi, the linear and
quadratic relations give similar results except for the most
metal-poor stars. 
\referee{Our calibration is presently quite uncertain near the
  metal-poor end, being based on one globular cluster (NGC\,4590).
  For this cluster, \citet{prit+2005} give a lower metallicity
  ([Fe/H]~$\sim -2.3$) than that adopted in Table~\ref{tab:paramGCS},
  yielding a better agreement with our linear calibration.}
%

%--- ERROR ON METALLICITY:

The main source of error on [Fe/H] (or [M/H]) 
is the uncertainty on the
measured equivalent width $\Sigma W$,
since other sources of error, such as photometric errors for stars on the
upper RGB of \leoi, the error on the HB level, or even the
uncertainties associated to the fit parameters of the calibration
relations, are negligible compared to the $\Sigma W$ measurement
errors.
%
%
% -Pag 7: contributo all'errore (commenti da MG):
%   M/H=0.395 W'-2.628
%   W'=EW+ \beta * V_HB
%   un errore dV_HB=0.1 sulla mag implica
%   dW'=\beta*dV_HB=0.627*dV_HB=0.06
%   d[MH]=0.395*dW'=0.02
%   da queste equazioni si ricava anche che l'errore sui parametri dei fit
%   e' trascurabile rispetto al contributo dovuto all'errore su EW
%   l'errore tipico dovuto a dEW e' dMH=0.14 dex
%
%
A metallicity uncertainty can be computed for each star by error
propagation using the values of $\epsilon\,_{\Sigma W}$ in
Table~\ref{t:data} and the calibrations in Table~\ref{tab:fit}.
However, a more meaningful metallicity uncertainty is obtained using
$\sigma_{W^\prime}$ as our estimate of the measurement error.  For the
quadratic [M/H] calibration in Table~\ref{tab:fit}, $\sigma_{W^\prime}
= 0.31$ \AA\ implies a [M/H] error $\sigma_{\rm err} \simeq 0.14$ dex
for stars with \mh~$=-1.20$.

%- INTRINSIC UNCERTAINTIES:
The sources of uncertainty
related to the CaT method itself
are more difficult to quantify and predict.
All traditional calibrations refer to Galactic
globular clusters, which are simple and nearly coeval old stellar
populations, and the applicability of these calibrations to
complex stellar populations is not obvious.
In our case, \leoi\ stars are on average several Gyr younger than
those in GCs. At a given luminosity, a star in \leoi\
has a different mass from a GC star with the same metallicity
and $V-\vhb$.
\referee{ 
  However, recent studies have shown that the CaT method can be used
  for complex stellar populations younger that those in globular
  clusters \citep{cole+2004,pont+2004,batt+2008}.
  \cite{batt+2008} have compared a linear metallicity calibration that
  uses the CaT of RGB stars in two dSph (Sculptor and Fornax) with
  spectroscopic [Fe/H] values obtained from their high-resolution studies.
  The metallicities are in good agreement, although with some residual
  trends of about 0.1--0.2 dex, in the range $-2.5 < \text{[Fe/H]} <
  -0.8$. These studies conclude that for ages older than 2.5 Gyr, the
  CaT line strengths are little affected by age, and suggest that the
  overall uncertainty related to age effects is $< 0.2$ dex. }

%--------------------------------------------
\subsection{A new metallicity scale?}

%--------------------------------------->
\begin{figure*}
\begin{center}
\includegraphics[width=.42\textwidth]{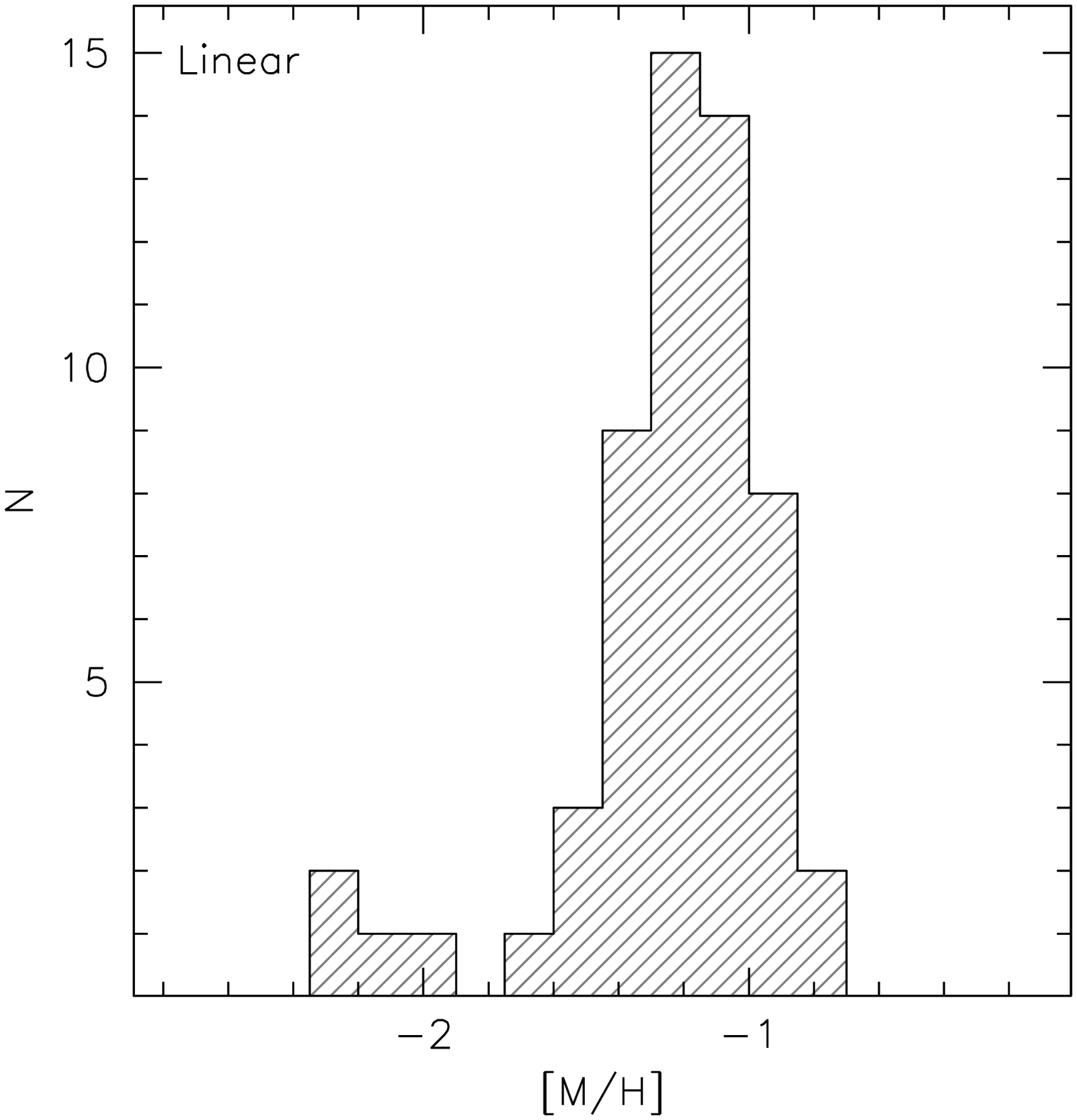}\hspace{.05\hsize}
\includegraphics[width=.42\textwidth]{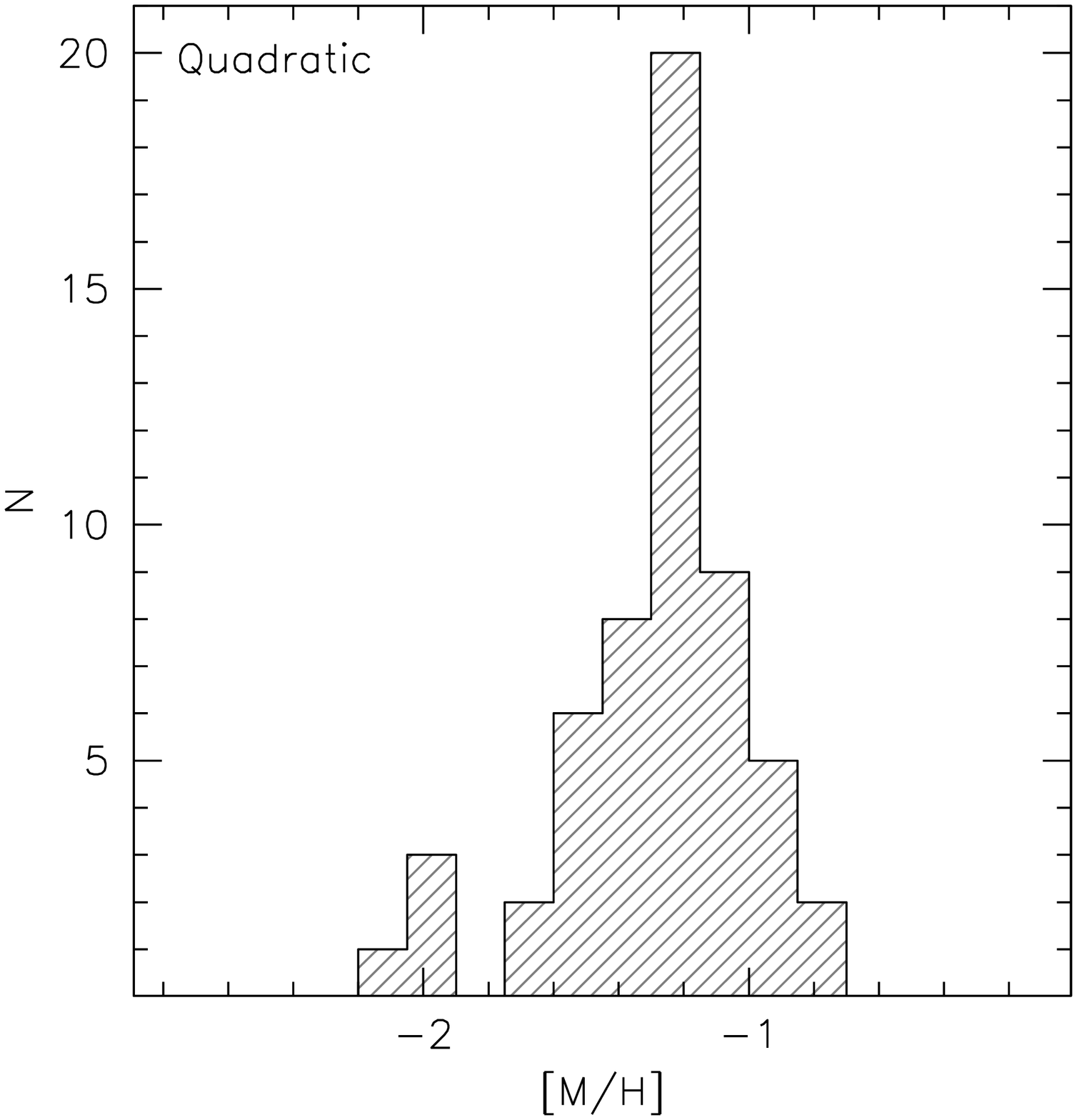}\\
\includegraphics[width=.42\textwidth]{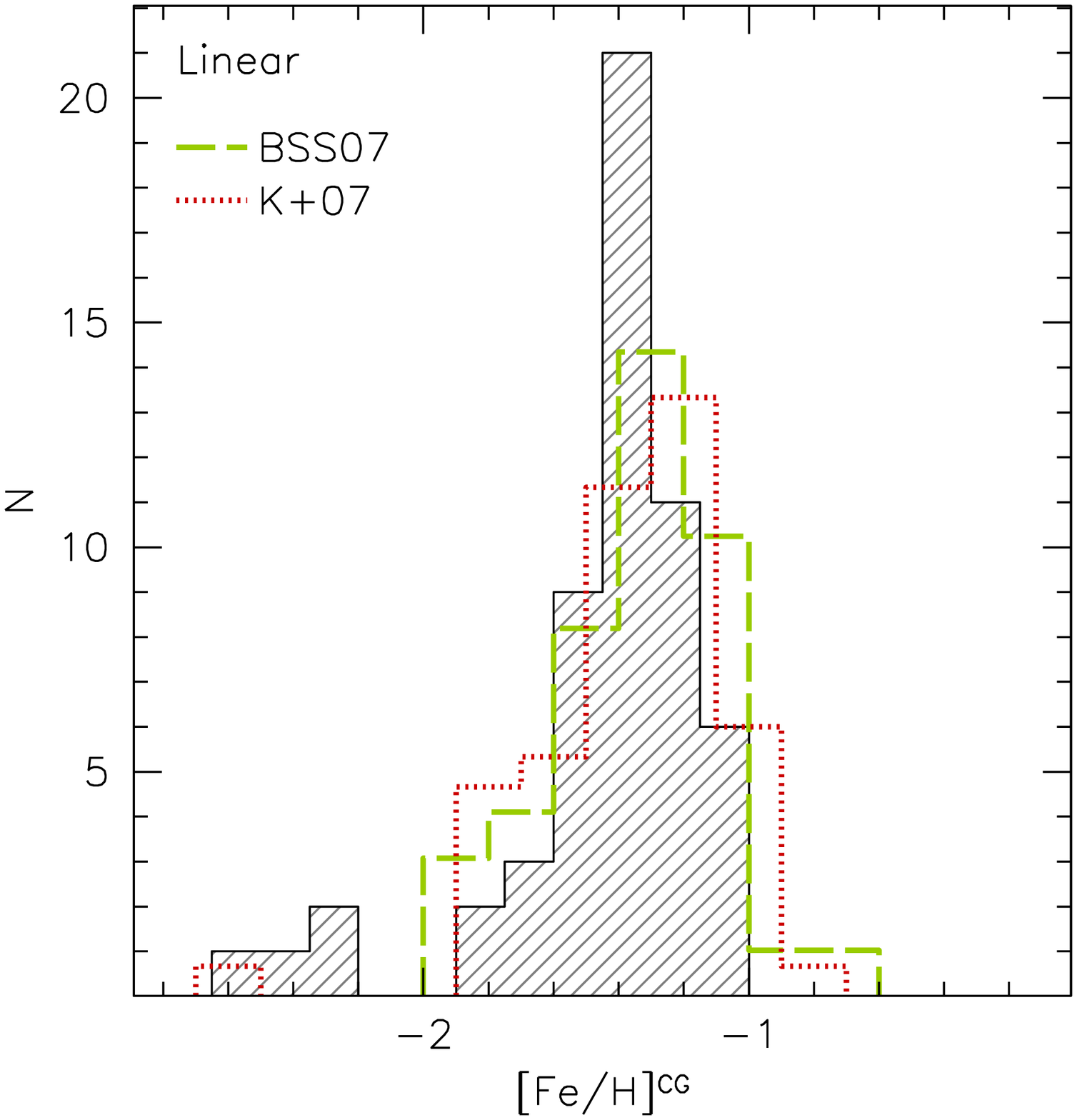}\hspace{.05\hsize}
\includegraphics[width=.42\textwidth]{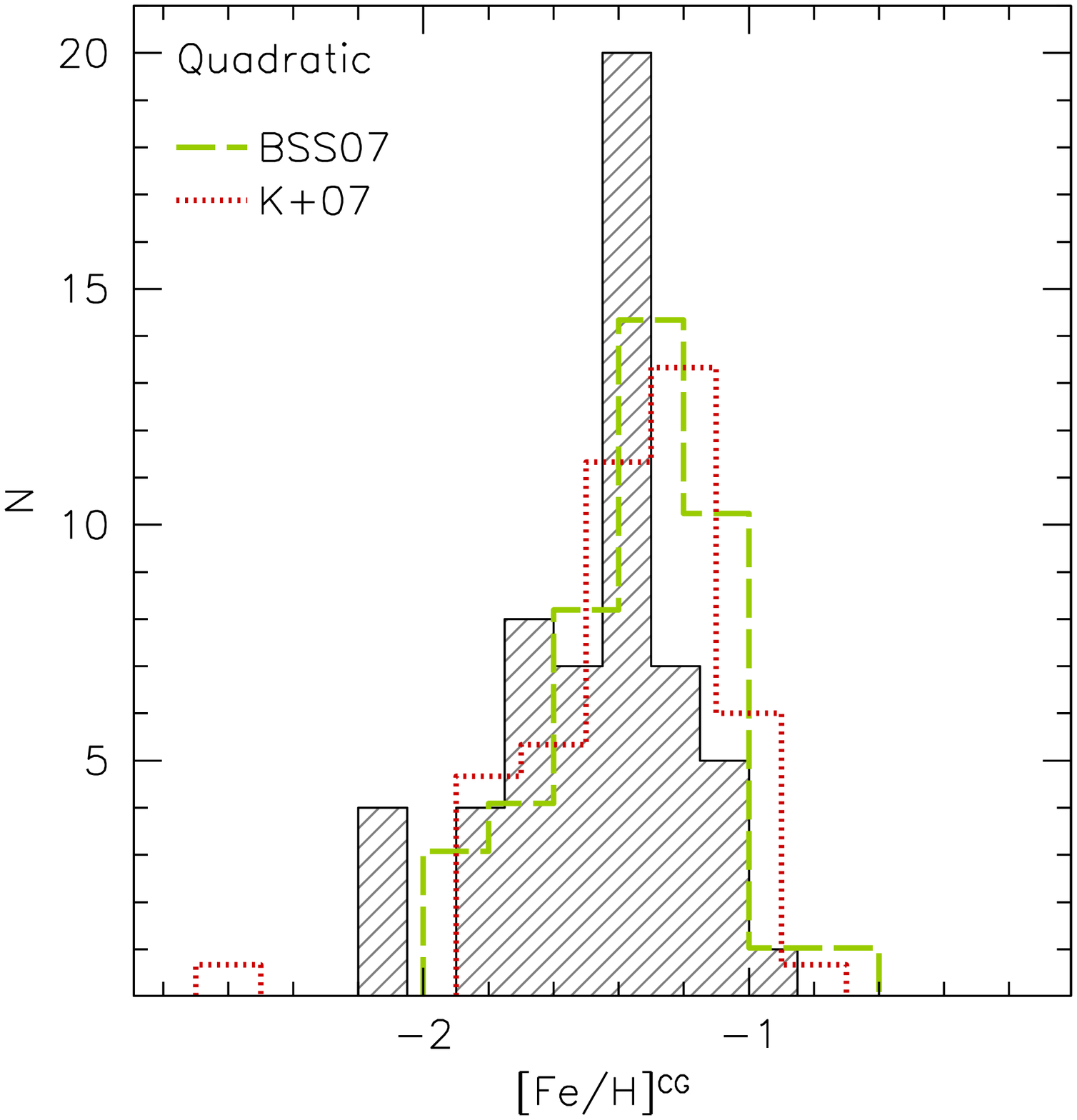}\\
\end{center}
\caption{Metallicity distribution of \leoi\ stars using a linear ({\it
left panels}) or quadratic ({\it right panels}) calibration relation.
In the {\it upper panels} we plot the MDF obtained with the \mh\
calibration. The metallicity distributions on the \feh\ scale of
\cite{carrgrat1997} are shown in the {\it lower panels}, together with
previous results from \citet{bosl+2007} and \cite{koch+2007leo1}.
\label{fig:his3p} }
\end{figure*}
%---------------------------------------|

  Most CaT metallicity measurements in nearby galaxies use a \feh\
  scale based on observations of Galactic globular
  clusters
%
%\referee{\footnote{
%
%    Recently, \cite{batt+2008} proposed a new calibration independent
%    of globular clusters, using high-resolution spectroscopic
%    metallicity of RGB stars in two dSph. Using their calibration for
%    our CaT measures is not feasible because we used a different
%    instrumental setup and used a slightly different method to
%    calculate the EW.}}
%
 \citep[e.g.][]{pont+2004,cole+2004,batt+2006,koch+2006}.
  However, the relative abundances of $\alpha$-elements (including Ca)
  in Local Group dwarfs are on average lower than in the
  Milky Way halo stars and GCs
  \citep{shet+2001, shet+2003, tols+2003, geis+2005,
    prit+2005}.
%
%- BOSLER APPROACH:
  To overcome this problem, \citet{bosl+2007} proposed a new
  calibration of CaT lines against the [Ca/H] abundance, based on
  high-resolution spectroscopy of Galactic star clusters.  
  \referee{
    In their hypothesis, the [Ca/H] calibration is less affected by
    the difference in [Ca/Fe] abundance ratios between red giant stars
    in globular clusters and dwarf spheroidal galaxies.}
  However, the strength of CaT lines is also determined by other
  parameters (gravity and $T_{\rm eff}$) in addition to Ca abundance.  
%
%-
  \referee{ A comparison of Ca abundances derived from CaT lines with
    the results of high-resolution abundance measurements for stars in
    two dSph galaxies \citep{batt+2008} shows that, while the CaT
    lines trace both Ca and Fe, their dependence on Fe abundance is
    stronger.
    Similarly, [Ca/H] ratios derived from CaT lines for stars in
    \leoii\ dSph and globular clusters \citep{shet+2009}
    systematically differ from those obtained from mid-resolution
    synthetic spectra, with a residual trend that is a function of
    metallicity.
  Since the effective temperature of red giants in globular clusters
  is driven by their global metallicity [M/H] \citep{sala+1993}, an
  empirical metallicity ranking based on CaT and a {\it global
    metallicity} [M/H] scale, proposed here for the first time,
  appears to be the most empirically sound.
In fact, \mh\ (or, equivalently, $Z$) takes into account the abundances
of both the $\alpha$-elements and Fe. Our data are therefore
calibrated using the \mh\ calibration in addition to the common \feh\
scales.
}

\subsection{A concluding remark}

\referee{
We conclude this section with a consideration that ought to be kept in
mind throughout all the following discussion.  While we give the
metallicity of the stars in three flavours (\feh$^{\rm ZW}$, \feh$^{\rm
  CG}$, and \mh), this does not imply that we are determining the three
metallicity parameters at the same time.
}
The only observable quantity is the reduced equivalent width
$W^\prime$.  The calibration of $W^\prime$ in terms of metallicity
relies on the assumption that $W^\prime$ is correlated with
metallicity, i.e. a star in Leo~I has the same metallicity as a star
in a GC with the same $W^\prime$.  The key questions are: what are the
real drivers that determine the strength of the CaT lines? Do two
stars with the same iron-to-hydrogen ratio but different
$\alpha$-elements composition have the same $W^\prime$?
\referee{Some of these effects have been discussed by
  \citet{batt+2008}, and are further addressed by a large
  observational program by our group whose results will be presented
  in future papers.}

% Answering this question is beyond the possibilities of the present
% study and would need a dedicated observing program, together with a
% careful comparison with results from high-resolution spectroscopy and
% synthetic spectra \referee{\citep[e.g.][]{batt+2008}}.

%%%%%%%%%%%%%%%%%%%%%%%%%%%%%%%%%%%%%%%%%%%%%%%%%%%%%%%%%%%%%%%%%%
\section{The metallicity of \leoi\  stars}
\label{s:mdf}

%-------------------------------------------------
\subsection{The observed metallicity distribution}

%----------------------------------------------->
\mytab{l r c c}
{Scale & fit & mean & $\sigma$ \\}
{
$ \feh^{\rm ZW}$&     linear& $-1.53$& $ 0.17$\\
$          $&  quadratic& $-1.55$& $ 0.21$\\
$ \feh^{\rm CG}$&     linear& $-1.37$& $ 0.18$\\
$          $&  quadratic& $-1.41$& $ 0.21$\\
$       \mh$&     linear& $-1.18$& $ 0.19$\\
$          $&  quadratic& $-1.22$& $ 0.20$\\
}
{Mean metallicity and standard deviation of red giants in
\leoi.}{tab:mdf}
%-----------------------------------------------|

  In Fig.~\ref{fig:his3p} we show the metallicity distribution of
  \leoi\ red giant stars as derived from our data using both the
  [Fe/H] metallicity scale of \citet{carrgrat1997} and the \mh\ scale.
  The parameters of the distribution (mean and standard deviation,
  excluding the 4 most metal-poor stars in Fig.~\ref{fig:his3p}) are
  given in Table \ref{tab:mdf} for both scales, along with the results
  on the \citet{zinnwest1984} scale for ease of comparison with
  previous literature.  The metallicities of individual \leoi\ stars
  are listed in Table~\ref{t:data}.  Using the CG97 scale, the
  distribution is centred at \feh~$\simeq -1.4$ with a standard
  deviation $\sigma_{\feh} \simeq 0.2$. For the \mh\ calibration, the
  average is \mh~$\simeq -1.2$ with the same scatter.

  The results obtained from the linear and quadratic calibrations are
  very similar in all cases, as expected since most of the \leoi\
  stars have metallicities lower than $\feh=-1.0$.  The choice of the
  linear or quadratic relation only affects the metallicity of 4
  metal-poor stars, having \mh$\lesssim-2$.  In this range, the
  calibration is extrapolated beyond the most metal-poor globular
  cluster, which makes the metallicity of the 4 stars quite uncertain
  and dependent on the adopted calibration.
  In the case of the CG97 calibration, which yields the lowest
  extrapolated values, the 4 stars have $-2.6 < {\rm [Fe/H]} < -2.2$.
  Visual inspection of the targets on a \leoi\ image indicates
  normal star-like profiles (i.e. no blends).
  We therefore conclude that, while a few stars may have low
  metallicity, there is so far no evidence of extremely metal-poor
  stars in \leoi.  Spectral synthesis methods
  \citep[see, e.g., ][]{ekirb+2008a} will be used in a future paper to
  obtain more secure metallicity estimates for these metal-poor stars
  from a different spectral interval.

  Our determination of the metallicity of \leoi\ agrees well with the
  results of \cite{bosl+2007} (${\rm [Fe/H]}=-1.34$) and
  \cite{koch+2007leo1} (\feh~$=-1.31$), in particular when a linear
  calibration is used as in the previous papers.  On the other hand,
  our \mh\ values are in better agreement with the [Ca/H] results of
  \cite{bosl+2007}.
\referee{
  Using 9 stars in common with \cite{bosl+2007} and 5 stars in common
  with \cite{koch+2007leo1}, we compared the metallicities
  star-by-star. The mean differences are 
  $\Delta \feh^{\rm CG}=-0.04 \pm 0.14$ (rms of the sample) and
  $\Delta \feh^{\rm CG}=-0.17 \pm 0.11$, respectively,
  in  agreement with the shifts between the mean values of the MDFs. The
  rms values are consistent with our measurement error (see next
  section).
}

%-----------------------
\subsection{Intrinsic metallicity dispersion and clues on the evolution of \leoi}
%\subsection{Instrinsic metallicity dispersion}

The observed MDF in Fig.~\ref{fig:his3p} is the convolution of the
intrinsic metallicity distribution of stars in \leoi\ and the
measurement errors.
The real abundance spread can be estimated by adopting a Gaussian
model for the intrinsic MDF, which yields
$\sigma_{\text{OBS}}^2=\sigma_{0}^2 +\sigma_\text{err}^2$, where
$\sigma_{\text{OBS}}$ is the observed metallicity dispersion,
$\sigma_{0}$ is the intrinsic dispersion, and
$\sigma_\text{err}$ is the measurement error.
 
If we adopt a quadratic [M/H] calibration and a typical measurement
error $\sigma_{\rm err} \simeq 0.14$ dex, the measurement scatter
largely contributes to the observed metallicity dispersion.  The
observed scatter implies an intrinsic metallicity dispersion
$\sigma_{\feh,0}=\fehdisptrue$ for the CG97 scale and
$\sigma_{\mh,0}=\mhdisptrue$ for [M/H].
The intrinsic abundance dispersion of \leoi\ stars is therefore very
small, even smaller than previously thought, and this happens
in spite of the relatively wide range of ages of the stellar
populations.  This is an important constraint to the chemical
evolution across the life of the galaxy.

\realfigure{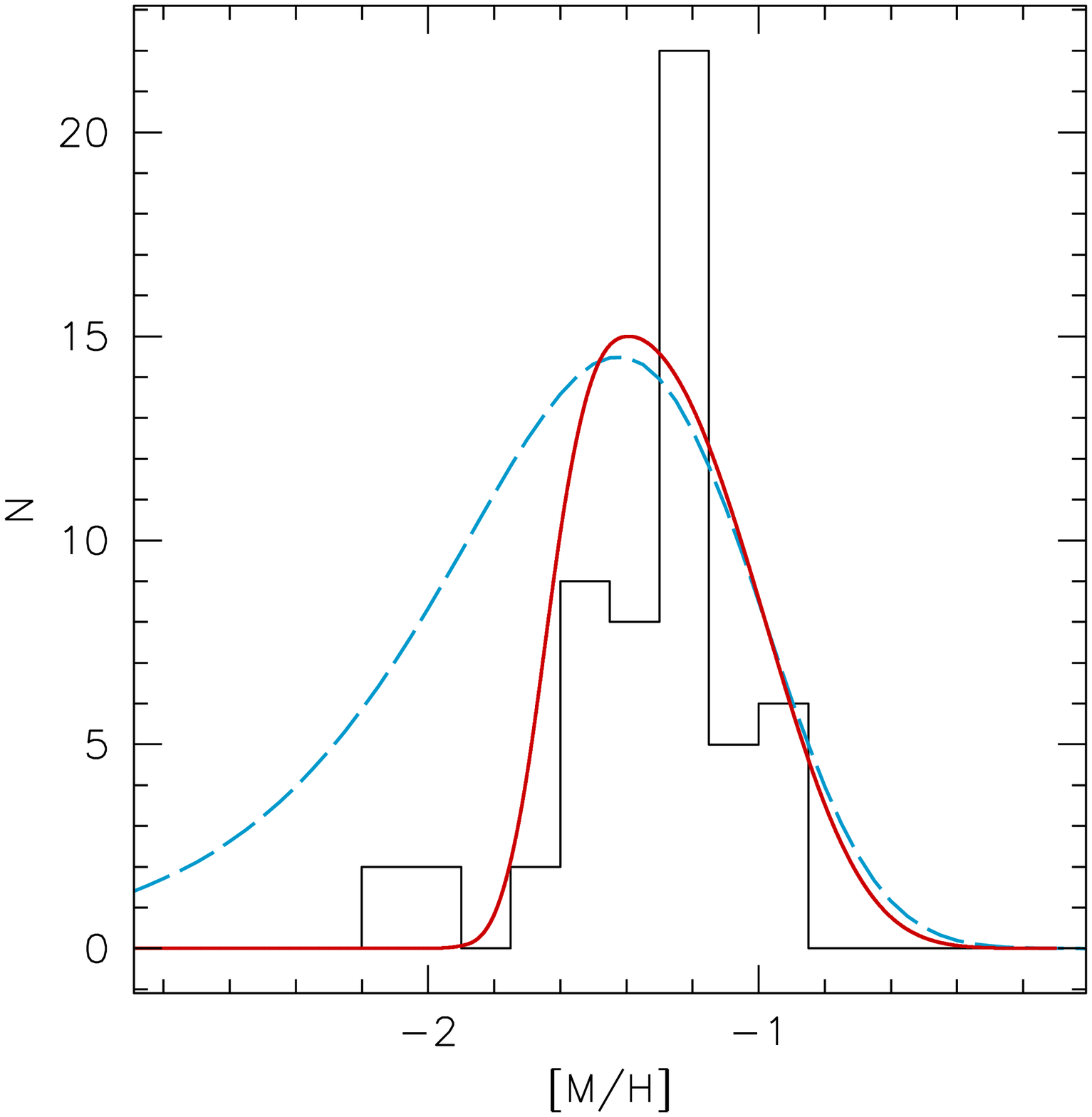} {The MDF of \leoi\ RGB stars on the \mh\
metallicity scale, compared with a simple model with a low effective
yield ({\it dashed line}) and a model with a prompt initial enrichment
({\it solid line}).  }{f:mdfok}
%-----------------------------------------------------|

In order to model the metallicity distribution and the chemical
evolution of \leoi, detailed models have to be put forth \citep[such as
those of ][]{lanfmatt2007} properly taking into account the chemical and
dynamical evolution of the galaxy.  However, some order-of-magnitude
physical insight on the evolution of \leoi can already be obtained using
basic considerations.
The metallicity distribution of RGB stars in \leoi\ is compared in
Fig.~\ref{f:mdfok} with a simple closed-box model with a low effective yield
consistent with a continuous loss of gas \citep[e.g.,][]{pagelbook}.
In order to reproduce the peak of the observed MDF, we have to adopt an
{\it effective} yield $y=0.025 Z_\odot$, and $y=0.040 Z_\odot$, for the
distributions based on the \feh\ and \mh\ metallicity scales,
respectively. This is clearly much lower than the value found in the
solar vicinity \citep[$y=1.2Z_\odot$; e.g.][]{port+2004}, in a way
consistent with the loss of metals driven by a galactic wind
\citep{hart1976,pagelbook}.
Still, even allowing for a gas outflow, the number of metal-poor stars
is largely overestimated by the simple model, as shown in Fig.
\ref{f:mdfok}. The fit is considerably improved by assuming a prompt
early enrichment with an initial metallicity $Z_0=0.02\, Z_\odot$ ($\feh
=-1.7$) (continuous line in Fig.~\ref{f:mdfok}).
Although very simplistic, this conclusion agrees with the
finding that the metal-poor tail of the MDF in 4 Local Group dwarf
spheroidal galaxies \citep{helm+2006} is significantly different from
that of the Galactic halo, lacking stars below \feh\ $=-3$.
What this ``toy model'' tells us is that the narrow MDF of the \leoi\
stars can be understood as a combination of fast enrichment from an
initial generation of stars, and subsequent loss of metals through
outflows.  This situation is common among Local Group dwarfs, but the
MDF of \leoi\ is the narrowest observed to date
\citep[cf. ][]{tols+2001,pont+2004,koch+2006,koch+2007leo2}.

%----------------------------------
\subsection{Radial metallicity gradients}

%----------------------------------------------------->
\realfigure{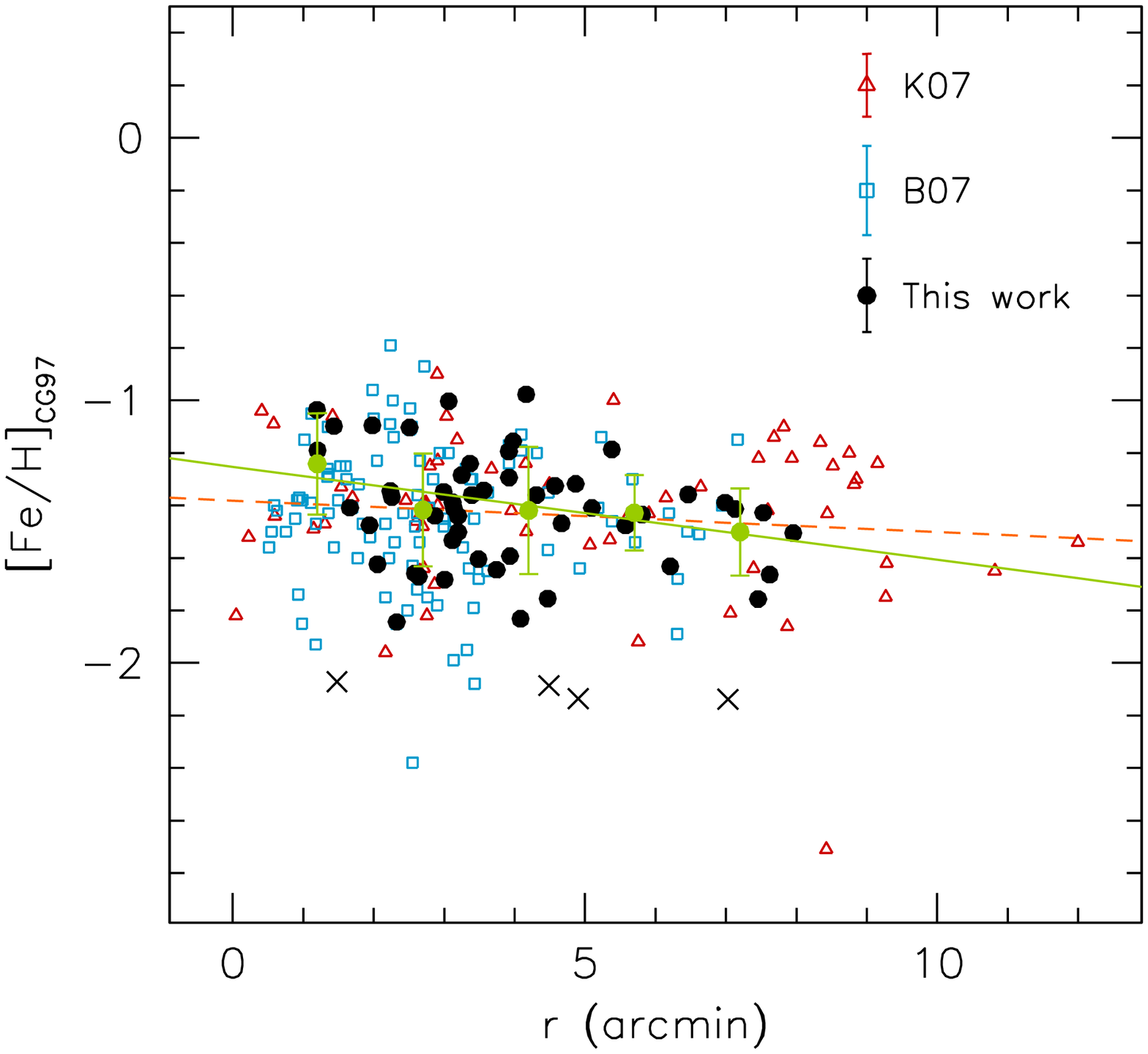}{  
Metallicities of \leoi\ stars on the [Fe/H]
  scale of CG97, plotted against the elliptical radius (see text).
  {\it Filled dots}: data in this paper; {\it open squares}: data from
  \citet{bosl+2007}; {\it open triangles}: \citet{koch+2007leo1}; {\it
    circles with error bars}: our data, binned in 1\farcm5 bins.  
The error bars of the binned data 
represent the abundance scatter ($1 \sigma$) in each bin.
The {\it crosses} are the 4 metal-poor stars in our sample.  
The typical errors of each study are shown in the upper right corner
of the plot. 
The {\it solid line} is a fit to our (binned) data, while the {\it
    dashed line} represents a fit to all available spectroscopic data.
}{fig:grad}
%-----------------------------------------------------|

Radial variations in the stellar populations are common in the dwarf
spheroidals of the Local Group, where the younger and more metal-rich
populations are often concentrated toward the galaxy centre
\citep{harb+2001,savi+2001, pont+2004,tols+2004,koch+2006}.  
\leoi\ remains one of the few dSph's showing little
evidence of a population gradient \citep{held+2000, koch+2007leo1}.  
In particular, the spectroscopic investigation of
\citet{koch+2007leo1}, extending to quite large radial distances, did 
not detect a significant metallicity gradient.

Our new spectroscopic sample of red giants allows us to further search
for radial variations in the metallicity of \leoi\ stars.  As the
radial coordinate, we have adopted the semi-major axis $r$ of ellipses
passing through the projected sky position of each star.  The ellipses
have the centre at \radec\  
and a fixed position angle and ellipticity
\citep[$\text{PA}=79^\circ$, $\epsilon=0.21$; ][]{irwihatz1995}.

In Fig.~\ref{fig:grad}, the metallicities of \leoi\ stars in our
sample (see Table~\ref{t:data}) are plotted against the distance from
the centre. 
To directly compare our data with results from previous studies, we
used the metallicity obtained from the calibration in terms of
[Fe/H]$^\text{CG}$. Data from \citet{bosl+2007} and
\citet{koch+2007leo1} were shifted to account for the small
differences in mean metallicity (of the order 0.1 dex or less) between
the MDF's (Fig.~\ref{fig:his3p}).  A linear relation was fitted both
to our data alone (solid line) and to all available metallicity
measurements in the literature (dashed line), excluding stars with
[Fe/H]~$< -2$.

The fit to our new FORS2 data yields a radial gradient 
of $-0.02$ dex arcmin$^{-1}$, or $\metgradkpc$ dex Kpc$^{-1}$.
%-dist.mod. = 22.04 -> 1' = 74.4 pc
%
In our sample, stars more metal-rich
than [Fe/H]~$= -1.3$ are only found in the central region of \leoi,
with $a < 5\farcm5$.
In contrast, the fit to the merged spectroscopic sample suggests a
radially constant metallicity, in agreement with the conclusions of 
\citet{koch+2007leo1}. 
To quantify the gradient, we have used a Kolmogorov-Smirnov test to
compare the metallicity distributions of stars with $1\farcm2 < a <
5\arcmin$ (inner sample) and $5\arcmin < a < 8\farcm2$ (outer sample),
considering only the radial interval covered by our data.  The two 
metallicity distributions are similar in shape, with the MDF in the
inner region peaked at higher metallicity.  The hypothesis that the
inner and outer sample are drawn from the same parent population can
be rejected at a 90\% level using our data, and only at a
non-significant 62\% level using all spectroscopic data.

We conclude that, while our data provide a hint of a weak radial
metallicity gradient in \leoi, the statistical significance of this
result is at present low. More stars need to be observed, particularly
in the outer region of the galaxy, before definite conclusions can be
drawn.

%%%%%%%%%%%%%%%%%%%%%%%%%%%%%%%%%%%%%%%%%%%%%%%%%%%%%%%%%%%%%%%%%%
\section{The age of  \leoi\ stars}\label{s:ages}

%----------------------------------------------------
\subsection{The age-metallicity relation}

\realfigure{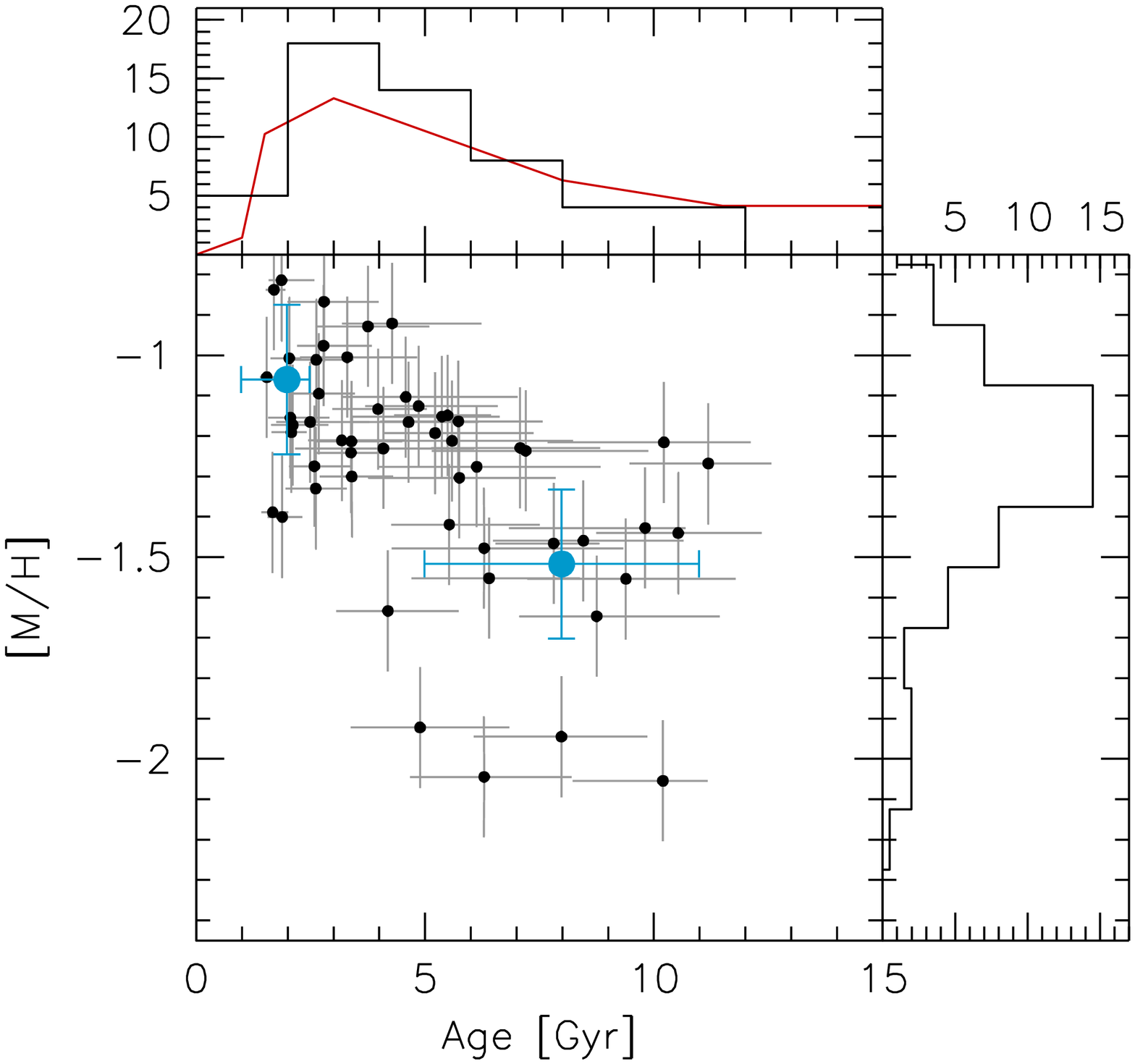}{
  The age-metallicity relation of \leoi\ RGB stars in our sample, on
  the \mh\ scale.  The error bars in age represent the first and third
  quartile of the confidence intervals obtained through Monte Carlo
  realisations (see text for details).  For the metallicity errors the
  representative value discussed in Sect.~\ref{s:metcalib} is adopted.  Also shown
  are the metallicities from high-resolution spectroscopy of two
  \leoi\ RGB stars from \cite{tols+2003} ({\it big filled circles}).
  The side histograms are the marginal distributions in metallicity
  and age. The {\it solid line} in the top panel represents the
  SFH derived by \cite{dolp2002} from HST
  photometry, normalised to the total number of stars in our sample.
}{fig:amr22}
%-----------------------------------------------------------|

With the stellar metallicities of \leoi\ stars known from
spectroscopy, ages could be estimated by comparing stars' locations in
the CMD with a grid of theoretical isochrones  \citep[the models of
][ were used to this purpose]{piet+2004}.
Absolute magnitudes and dereddened colours were computed adopting a
colour excess $E_{B-V} = 0.04$, a total-to-differential extinction
ratio $R_{V} = A_{V} / E_{B-V} = 3.1$, and a true distance modulus
$(m-M)_0 = 22.04$ \nirpapercitep.

The stellar ages were interpolated in two steps.  First, we used a set
of theoretical isochrones of fixed age and different metallicities to
find, for a star of given age and luminosity, a metallicity-colour
relation and (from the known colour) an interpolated metallicity.
This step was repeated for all model ages, yielding for each data
point in the CMD a set of theoretical age-metallicity pairs each 
consistent with the star's magnitude and colour.
This age-metallicity look-up table (spanning the full range from old,
metal-poor stars to young, metal-rich stars)
allowed us to compute an interpolated age for each star from its
spectroscopic metallicity. We chose the [M/H] scale for the input
value, as the most directly related to the mass fraction of 
metals ($Z$) used in stellar models.  For a number of stars, ages
could not be derived because the observed colour and/or magnitude were 
outside the range covered by the isochrones.

The method was checked against Galactic globular clusters with ages
given by the literature.  In particular, a small correction was
applied to the isochrone colours so as to yield a correct age for
NGC\,5904 (M\,5), assumed to be 12 Gyr \citep{sand+1996} and the closest in
metallicity to \leoi\ among the clusters listed in
Table~\ref{tab:paramGCS}. Thus, our ages for \leoi\ stars are
essentially referred to M\,5, which is in our view the most correct
approach given the considerable uncertainties in the isochrone colours.

The resulting ages are listed in Table~\ref{t:data} along with their
uncertainties, estimated as follows.  For each star, we performed a
set of \numexp\ Monte Carlo experiments by randomly varying the input
quantities in intervals consistent with their uncertainties. We
adopted a standard error 0.02 mag in the $V$ magnitude, 0.05 mag in
$(B-V)$ colour, and 0.15 dex in [M/H]. The latter was chosen
conservatively large to account for the inherent uncertainties in the
metallicity scale.
The median and quartiles of the age distributions of randomly
generated ``stars'' corresponding to each observed stars are listed in
Table~\ref{t:data}. In general, the uncertainties are of the order
50\%, which reasonably reflects the large uncertainties in the
process.

The age-metallicity relation derived from our data is shown in
Fig.~\ref{fig:amr22}.  The chemical evolution of \leoi\ seems to be
very slow, in accord with the narrowness of the MDF.  If the 4
metal-poor stars are excluded, there is a trend for stars younger than
5 Gyr to be on average more metal-rich by about 0.2--0.3 dex.  Similar
conclusions were drawn, from a different data set, by \citet{bosl+2004}.
The scatter in the age-metallicity relation appears to be smaller than
observed in other galaxies \citep[][and references
therein]{batt+2006,tols+2003}.
Our data are in agreement with the results from high-resolution
spectroscopy for 2 stars \citep{tols+2003}.

\marco{
We can use our age determinations also to obtain a SFH of \leoi.
This can be done since our target selection was designed to avoid
any bias in age and/or metallicity (see Sect. \ref{s:obsredu});
the age distribution of our target stars is then proportional to the SFH.
Our age measures are in agreement with the SFHs derived by HST
photometry.  In the upper panel of Fig. \ref{fig:amr22} our age
distribution is compared with the \cite{dolp2002} SFH, showing only
small differences that can be explained by statistical fluctuations.
}

%-------------------------
\subsection{Radial distribution of stellar ages}

To complete our analysis of the population gradients, we investigated
the possible presence of a radial variation in the age of \leoi\ stars.
Figure~\ref{fig:agegrad} shows the ages plotted against the elliptical
distance.
While old stars are found at all radii, young stars appear to be
concentrated at small distances from the centre.  To quantify this
finding, we have plotted in Fig.~\ref{fig:agegrad} the cumulative
distributions of two subsamples of stars in the inner ($r < 3\farcm9$)
and outer ($r > 3\farcm9$) region, respectively.  This limit was
chosen to have the same number of stars in each subsample.  A
Kolmogorov-Smirnov test indicates that the null hypothesis that the
distributions are drawn from the same parent population can be
rejected at $ > 99.9$\% confidence level.  Similarly, we have
plotted the cumulative distributions of two subsamples of stars with
age $< 4.6$ Gyr and $> 4.6$ Gyr. Also in this case, the probability of
the null hypothesis can be rejected at a level $> 99.9$\%.

This suggestion of an age gradient among RGB stars is strengthened by
the detection of a radial gradient in the fraction of upper-AGB stars
in \leoi, which points to a concentration of intermediate-age
populations towards the galaxy centre \nirpapercitep.

%---------------------------------------------------------------->
\realfigure{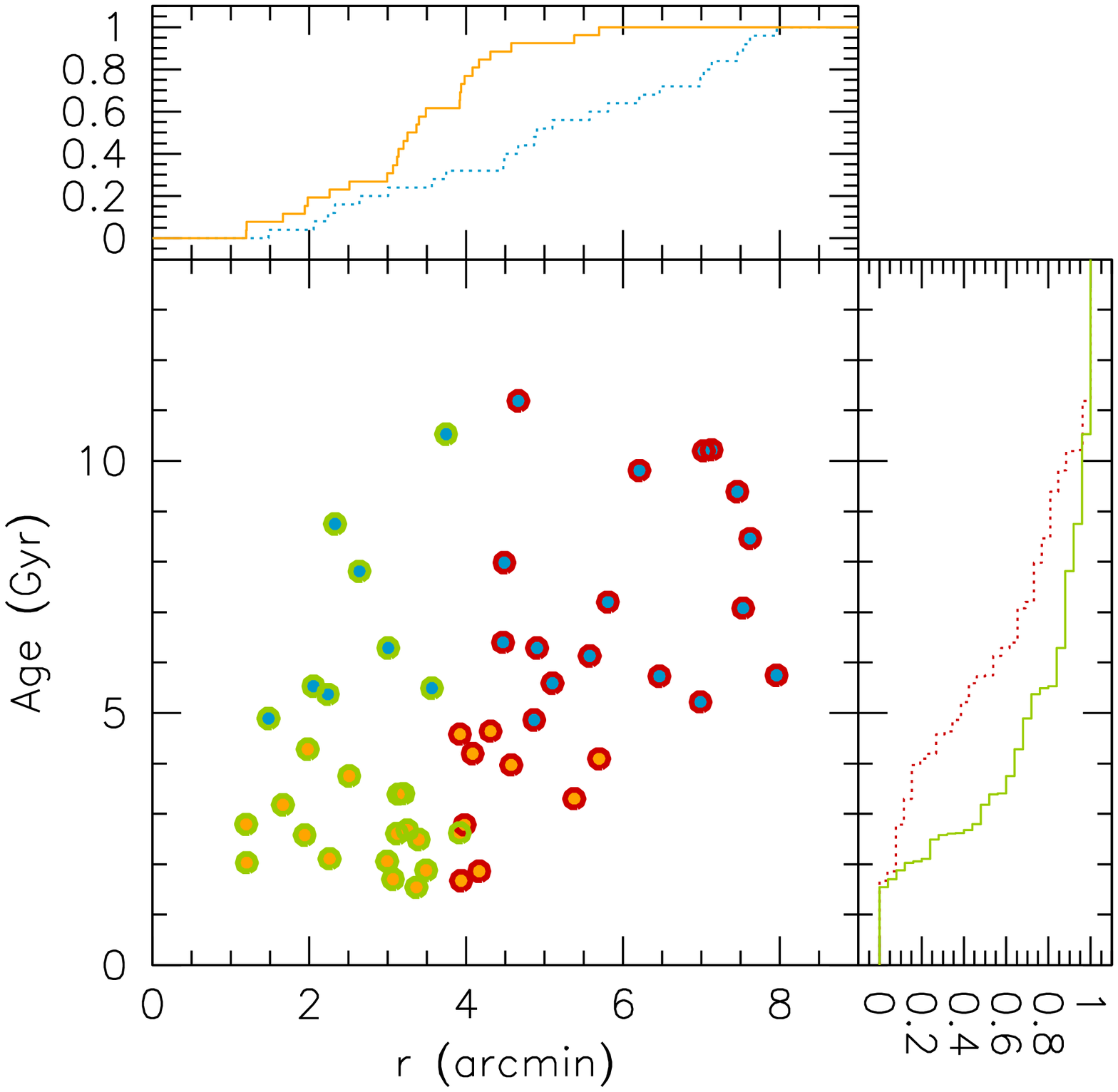}{
  Ages of the \leoi\ stars in our sample as a function of the
  elliptical radius ({\it central panel}).  The {\it upper panel}
  shows the cumulative radial distributions of stars with ages smaller
  and larger (dotted line) than 4.6 Gyr. The cumulative age
  distributions of stars in the inner ($r < 3\farcm9$) and outer ($r >
  3\farcm9$, dotted line) region are shown in the {\it right panel}
  (the different age and radial intervals are colour-coded in the
  electronic version of the journal).  Younger stars appear to be more
  concentrated towards the centre of \leoi.
}{fig:agegrad}
%----------------------------------------------------------------|

%%%%%%%%%%%%%%%%%%%%%%%%%%%%%%%%%%%%%%%%%%%%%%%%%%%%%%%%%%%%%%%%%%

%----------------------------------------------------------------
\section{Summary and conclusions}
\label{s:sum}

  We have presented spectroscopic measurements of RGB stars in the
  \leoi\ dSph from observations carried out with the FORS2 
  spectrograph at the ESO VLT.  We derived radial velocities for 57
  stars with good S/N ratio, 54 of which have been found to be \leoi\
  members. Among these, 14 stars are in common with previous
  spectroscopic studies.

%-METALL. MEASUREMENTS:
  We measured the metallicities of RGB stars in \leoi\ from the
  equivalent widths of \caii\ triplet lines, using the \feh\
  metallicity scales of \citet{carrgrat1997} and \citet{zinnwest1984}.
  In addition, we derived a new calibration tied to the \mh\ ranking
  of Galactic globular clusters, which accounts for the abundance of
  both Fe-group and $\alpha$ elements.

%- NARROW MDF:
  The metallicity distribution (MDF) of \leoi\ stars is symmetric and
  very narrow.  If we adopt a quadratic calibration of \caii\ line
  strengths against [Fe/H], the mean metallicity is $\feh=-1.41$ with
  a measured dispersion $0.21$ dex on the \cite{carrgrat1997} scale,
  in agreement with previous spectroscopic studies.  The new \mh\
  calibration yields a mean value \mh\ $=-1.22$ with a dispersion
  $0.20$ dex. By subtracting the measurement errors, we
  estimated a very low intrinsic metallicity dispersion, $\sigma_{\rm
    [M/H]} = \mhdisptrue$, which represents a constraint for 
  modelling the chemical evolution of this isolated dwarf galaxy.
  As pointed out by previous studies, this narrow MDF is inconsistent
  with a simple ``closed-box'' chemical evolution model, even adopting
  a very low effective yield to account for galactic outflows 
  expelling the metals produced by SNe winds. A prompt initial chemical
  enrichment may explain the very small number of extremely metal poor
  stars (we find only 4 stars with [Fe/H]~$< -2$).  Together, the two
  effects can explain the small abundance dispersion of \leoi\ stars,
  which gives the narrowest observed MDF among Local Group dwarf
  galaxies. However, detailed chemical evolution models \citep[e.g.,
  ][]{lanfmatt2007} are needed to gain a complete picture of the
  evolution of \leoi.

%-METALLICITY GRADIENT:
  Our data for RGB stars also provide an indication of a weak radial
  metallicity gradient in \leoi, of $\metgradkpc$ dex Kpc$^{-1}$.
  In fact, all of our stars with \mh$>-1.3$ are found in the inner region
  ($r\lesssim 5^\prime$).
  However, by combining our observations with previous spectroscopic
  datasets in the literature, the radial variation becomes
  insignificant. 
 \referee{
   More observations in the
   outskirt of \leoi\ with a quality comparable to those presented here,
   are required to definitively establish the presence of an abundance
   gradient.} 
%
% \referee{
% %
%   Our spectra have an higher S/N ratio than those used in previous
%   works, so our hint for a radial trend must be taken into account.
%   Thus, at present, the evidence for an abundance gradient is not
%   conclusive, but this work shows that more observations in the
%   outskirt of \leoi\ with a quality comparable to those presented here,
%   are required to clear up any doubt on this point.
% %
% } 

%Thus, at present, the evidence for an abundance gradient is not
%conclusive.

%- AGES and AMR:
  The metallicities of the RGB stars in our \leoi\ sample have been
  combined with existing photometric data to yield age estimates and
  an age-metallicity relation.
  Our age determinations are
  consistent with the SFH derived by \cite{gall+1999}
  and \cite{dolp2002} from HST photometry.
  The age-metallicity relation of \leoi\
  red giants is quite flat, again suggesting a rapid initial
  enrichment.  An increase in metal
  abundance by $\sim 0.2-0.3$ dex in the last 5~Gyr is possibly
  related to the main star-formation episode at intermediate ages.
  Since \leoi\ only hosts a minor old ($>10$ Gyr) stellar component,
  the chemical history of the galaxy is not well constrained at early
  epochs.  
  Its most metal-poor stars must have formed out of a medium
  pre-enriched by a lost generation of stars, either before or after
  the galaxy had started assembling.
  % 

%- AGE GRADIENT:
  We have provided the first evidence of a radial variation in the
  ages of red giants in \leoi.  Despite the uncertainties in age
  determination, our direct measurement of a radial variation of
  stellar ages seems quite convincing, with a Kormogorov-Smirnov test
  confirming, at a high level of statistical significance, that stars
  in the inner part of \leoi\ are on average younger than those in the
  outer regions.  This result agrees with the conclusions of a
  parallel study of intermediate-age AGB stars in \leoi\ from
  near-infrared photometry \nirpapercitep.
  In the emerging scenario, the first generation of \leoi\ stars
  uniformly formed throughout this isolated dwarf spheroidal galaxy.
  The bulk of intermediate-age stars originated from an interstellar
  medium, poorly enriched by previous stellar generations mainly
  because of the effects of stellar winds.
  Younger stellar populations preferentially formed in the central
  regions, from gas somewhat enriched as seen from the age-metallicity relation
  in the last few Gyr.
  In this framework, our results on the radial distribution of \leoi\
  stellar populations are not in contrast with previous results which
  found no gradients.
  The lack of detection of an age gradient by \citet{held+2000} can be
  explained considering that the mean age of the red-clump stars used by
  \cite{held+2000} as tracers of intermediate-age populations, 
  is $\sim 5$~Gyr, which is older than
  that of upper AGB stars used by \citet{held+2008leo1}.  As shown in Fig. \ref{fig:agegrad}, there
  are no clear radial variations in the age distribution of \leoi\
  stars with ages greater than $\sim 5$~Gyr.

%-----------------------------------------------------------------
\begin{acknowledgements}
  We thank A. Koch for providing us with unpublished data.  M.G.\
  wishes to thank the European Southern Observatory at Santiago, Chile
  for partial funding through DGDF and for hospitality during a visit
  in which this paper was partially written.  This research was
  partially funded by PRIN MIUR 2007 ``Galactic astroarchaeology: the
  local route to cosmology'' (P.I. F. Matteucci).
\end{acknowledgements}

\bibliographystyle{aa}
\bibliography{1578}

%- uncomment to omit tables:
%\end{document}

%----------------------------- LONGTABLES ---------------------------
%\renewcommand{\baselinestretch}{1.}) 

\onecolumn
\clearpage
\setcounter{table}{0}

\mytabnote{%put the column formatting -- r@{\hspace{10pt}}
r c c c c r r@{\hspace{15pt}}
c l
}{%put the tab header
\multicolumn{1}{c}{ID}&
\multicolumn{1}{c}{$\alpha$ (J2000)}&
\multicolumn{1}{c}{$\delta$ (J2000)}&
\multicolumn{1}{c}{$B-V$}&
\multicolumn{1}{c}{$V$}&
\multicolumn{1}{c}{$v$}&
\multicolumn{1}{c}{$\Delta v$}&
\multicolumn{1}{c}{S/N}&
\multicolumn{1}{l}{other}
\\
&
&
&
&
&
\multicolumn{1}{c}{\kms}&
\multicolumn{1}{c}{\kms}&
&
\\
}{%put the tab body
 %ID &            RA &           DEC &      BV &       V &      Vel &   dVel   & S/N&comm. \\  %   ID cat orig
   1 &   10:08:28.43 &   +12:15:54.1 &    1.39 &   19.62 & $ 268.2$ &    1.5   & 57 &      \\ %    23  r 36436
   2 &   10:08:16.47 &   +12:15:59.5 &    1.48 &   19.59 & $ 264.5$ &    3.2   & 46 &      \\ %    18  r 24984
   3 &   10:08:36.41 &   +12:16:02.2 &    1.46 &   19.89 & $ 261.9$ &    5.7   & 37 &      \\ %    28  r 25309
   4 &   10:08:17.58 &   +12:16:15.7 &    1.61 &   19.55 & $ 267.0$ &    0.9   & 42 &      \\ %    19  r 23539
   5 &   10:08:19.45 &   +12:16:34.1 &    1.10 &   20.28 & $ 268.6$ &    3.2   & 32 & B8391  \\ %    20  r 20964
   6 &   10:08:40.46 &   +12:16:38.0 &    1.35 &   19.78 & $ 273.5$ &    0.4   & 36 & B25113 \\ %    22  r 19866
   7 &   10:08:31.00 &   +12:16:40.6 &    1.32 &   19.85 & $ 282.4$ &    3.1   & 58 & B18214 \\ %    25  r 32810
   8 &   10:08:03.62 &   +12:16:46.4 &    1.50 &   19.74 & $ 301.0$ &    7.4   & 59 &      \\ %    36  r 40728
   9 &   10:08:32.56 &   +12:16:54.9 &    1.19 &   20.45 & $ 290.7$ &    1.5   & 33 &      \\ %    26  r 30630
  10 &   10:08:08.79 &   +12:17:05.9 &    1.07 &   20.00 & $ 304.3$ &    2.2   & 90 &      \\ %    40  r 34557
  11 &   10:08:10.61 &   +12:17:08.0 &    1.20 &   20.05 & $ 262.2$ &    6.8   & 62 & B4173 \\ %    41  r 32361
  12 &   10:08:49.42 &   +12:17:14.8 &    1.18 &   20.44 & $ 281.2$ &    2.5   & 18 &      \\ %     2 rw 200980
  13 &   10:08:34.90 &   +12:17:17.8 &    1.16 &   20.01 & $ 287.8$ &    6.8   & 38 &      \\ %    27  r 27421
  14 &   10:08:39.66 &   +12:17:19.9 &    1.37 &   19.68 & $ 258.4$ &    3.7   & 44 & K195 \\ %     3  r 20953
  15 &   10:08:57.38 &   +12:17:20.8 &    1.18 &   20.44 & $ 280.3$ &    6.6   & 34 & K833 \\ %    15 rw 201004
  16 &   10:08:55.47 &   +12:17:21.9 &    1.26 &   20.14 & $ 274.2$ &   10.9   & 43 &      \\ %    13 rw 201012
  17 &   10:08:11.82 &   +12:17:29.4 &    1.05 &   20.31 & $ 260.0$ &    0.6   & 51 &      \\ %    42  r 30858
  18 &   10:08:29.89 &   +12:17:31.7 &    1.30 &   19.67 & $ 262.8$ &    4.8   & 43 &      \\ %    24  r 34381
  19 &   10:08:07.52 &   +12:17:34.6 &    1.25 &   20.16 & $ 269.6$ &    7.4   & 52 &      \\ %    39  r 36120
  20 &   10:08:14.21 &   +12:17:36.1 &    1.11 &   19.90 & $ 247.6$ &   32.2   & 67 & B5496 \\ %    34  r 27851
  21 &   10:08:15.40 &   +12:17:38.3 &    1.43 &   19.48 & $ 266.1$ &    3.0   & 37 &      \\ %    17  r 26320
  22 &   10:08:46.36 &   +12:17:41.3 &    1.33 &   20.32 & $ 261.9$ &    2.9   & 30 &      \\ %     8  r 12465
  23 &   10:08:20.50 &   +12:17:45.0 &    1.61 &   19.64 & $ 263.7$ &    0.3   & 50 &      \\ %    21  r 19530
  24 &   10:08:06.29 &   +12:17:45.3 &    1.28 &   20.16 & $ 269.9$ &    7.1   & 49 & B3135 \\ %    38  r 37584
  25 &   10:08:50.70 &   +12:17:46.9 &    1.15 &   20.31 & $ 270.5$ &    6.7   & 35 & K677 \\ %    10 rw 201168
  26 &   10:07:55.77 &   +12:17:55.2 &    1.14 &   20.48 & $ 277.3$ &    6.0   & 47 &      \\ %    30 rw 201258
  27 &   10:08:01.84 &   +12:17:56.6 &    1.43 &   19.82 & $ 275.4$ &    5.4   & 61 & B2488 \\ %    33 rw 201269
  28 &   10:08:45.25 &   +12:17:57.8 &    1.28 &   20.08 & $ 248.9$ &    5.5   & 52 &      \\ %     7  r 13814
  29 &   10:08:38.58 &   +12:18:21.0 &    0.94 &   19.71 & $ 274.5$ &    8.3   & 41 &      \\ %     1  r 25974
  30 &   10:08:56.58 &   +12:18:22.4 &    1.13 &   20.22 & $ 271.1$ &    0.3   & 30 &      \\ %    14 rw 201382
  31 &   10:08:34.85 &   +12:18:22.6 &    1.24 &   19.98 & $ 256.2$ &    0.1   & 47 &      \\ %    58  r 31733
  32 &   10:08:15.96 &   +12:18:25.3 &    1.17 &   20.25 & $ 282.9$ &    3.0   & 34 &      \\ %    46  r 23503
  33 &   10:07:57.28 &   +12:18:26.1 &    1.40 &   19.74 & $ 284.3$ &    7.5   & 69 &      \\ %    31 rw 201416
  34 &   10:08:44.09 &   +12:18:29.0 &    1.13 &   19.66 & $ 278.8$ &    9.5   & 42 &      \\ %     6  r 17954
  35 &   10:08:47.87 &   +12:18:29.9 &    1.46 &   19.70 & $ 248.8$ &    8.0   & 39 &      \\ %     9  r 12705
  36 &   10:08:36.01 &   +12:18:32.6 &    1.30 &   19.58 & $ 254.0$ &    0.6   & 45 &      \\ %    59  r 29925
  37 &   10:07:59.36 &   +12:18:35.7 &    1.60 &   19.75 & $ 273.9$ &    6.3   & 54 &      \\ %    32 rw 201470
  38 &   10:08:58.70 &   +12:18:37.1 &    1.44 &   19.85 & $ 257.2$ &    8.2   & 46 &      \\ %    16 rw 201479
  39 &   10:08:41.92 &   +12:18:39.7 &    1.16 &   20.22 & $ 250.4$ &    7.8   & 31 & B25820 \\ %     5  r 21033
  40 &   10:08:51.73 &   +12:18:45.7 &    1.28 &   20.23 & $ 278.9$ &    3.0   & 49 &      \\ %    11 rw 201535
  41 &   10:08:20.21 &   +12:18:46.6 &    1.41 &   19.62 & $ 258.0$ &    0.1   & 27 &      \\ %    49  r 18263
  42 &   10:08:33.75 &   +12:18:47.0 &    1.26 &   19.23 & $ 275.4$ &    0.0   & 50 &      \\ %    57  r 33719
  43 &   10:08:24.22 &   +12:18:53.3 &    1.38 &   19.93 & $ 259.0$ &    2.4   & 42 &      \\ %    44  r 13082
  44 &   10:08:40.60 &   +12:19:02.8 &    1.25 &   19.66 & $ 276.9$ &    2.7   & 48 &      \\ %    54  r 22971
  45 &   10:08:30.67 &   +12:19:30.0 &    1.00 &   20.34 & $ 259.4$ &    4.6   & 39 &      \\ %    56  r 38168
  46 &   10:08:21.15 &   +12:19:43.6 &    1.52 &   19.72 & $ 270.8$ &    1.0   & 35 &      \\ %    50  r 17061
  47 &   10:08:39.42 &   +12:20:05.9 &    1.16 &   19.74 & $ 271.2$ &    3.8   & 51 &      \\ %    53  r 24744
  48 &   10:08:37.29 &   +12:20:12.2 &    1.24 &   19.88 & $ 270.7$ &    1.3   & 43 & K351 \\ %    52  r 27889
  49 &   10:08:13.35 &   +12:20:13.8 &    1.27 &   19.64 & $ 292.2$ &    2.9   & 55 &      \\ %    43  r 26633
  50 &   10:08:22.17 &   +12:20:14.9 &    1.38 &   19.15 & $ 289.8$ &    0.9   & 59 &      \\ %    51  r 15860
  51 &   10:08:28.41 &   +12:20:29.6 &    1.51 &   19.61 & $ 267.6$ &    2.1   & 64 & K137 \\ %    55  r 41546
  52 &   10:08:14.96 &   +12:20:43.9 &    1.47 &   19.58 & $ 266.6$ &    2.9   & 50 &      \\ %    45  r 24711
  53 &   10:08:19.17 &   +12:20:48.9 &    1.38 &   19.87 & $ 255.3$ &    2.0   & 38 & B8203 \\ %    48  r 19509
  54 &   10:08:18.24 &   +12:20:54.6 &    1.27 &   20.02 & $ 268.0$ &    0.7   & 31 &      \\ %    47  r 20668
 101 &   10:07:51.18 &   +12:17:36.6 &    1.52 &   19.56 & $  49.6$ &    8.5   & 72 &      \\ %    29 rw 201151
 102 &   10:08:05.23 &   +12:18:13.5 &    1.05 &   19.96 & $ -27.9$ &    5.5   & 52 &      \\ %    37  r 36036
 103 &   10:08:53.43 &   +12:18:27.4 &    1.36 &   19.88 & $  90.4$ &    8.7   & 55 &      \\ %    12 rw 201417

}
{
%put the caption
Spectroscopic sample in the \leoi\ field.   
}{%put the label
t:data2}{%size
normalsize}{{\it Notes}. $v$ is the heliocentric radial velocity and
$\Delta v$ the absolute semi-difference in radial velocity of the
individual spectra. The last column gives the identification of stars
in common with \citet{bosl+2007} and \citet{koch+2007leo1}.}

%-----------------------------------------------------------------------
\clearpage
\setcounter{table}{2} %- force to ``Table 3''
%------------------------------------------>
%%%%   l r r c c c r@{\hspace{10pt}} r@{\hspace{10pt}}

\mytabnote{%put the column formatting
r  c c c 
c c c 
r r c
}{%put the tab header
ID& 
$\Sigma W$& 
$\epsilon\,_{\Sigma W}$& 
$W^\prime$&
\feh$^{\rm ZW}$& 
\feh$^{\rm CG}$&
\mh& 
age&
$\Delta$age$^-$& 
$\Delta$age$^+$ 
\\
&
(\AA)&
(\AA)&
(\AA)&
& & &
(Gyr)&
(Gyr)&
(Gyr)\\
}{%put the tab body
 %ID &      W &     dW &     Wp & FeH_ZW & FeH_CG &     MH &    Age &  dAge- &  dAge+ \\ %   ID cat orig
   1 &   5.58 &   0.11 &   3.71 & $ -1.55$ & $ -1.41$ & $ -1.21$ &      3.4 &      0.8 &      1.1 \\ %   23  r 36436
   2 &   6.12 &   0.38 &   4.23 & $ -1.28$ & $ -1.16$ & $ -0.98$ &      2.8 &      0.6 &      1.1 \\ %   18  r 24984
   3 &   4.85 &   0.29 &   3.16 & $ -1.78$ & $ -1.64$ & $ -1.44$ &     10.5 &      1.8 &      1.8 \\ %   28  r 25309
   4 &   5.77 &   0.04 &   3.86 & $ -1.48$ & $ -1.34$ & $ -1.15$ &      5.5 &      1.2 &      0.9 \\ %   19  r 23539
   5 &   5.30 &   0.19 &   3.85 & $ -1.48$ & $ -1.35$ & $ -1.15$ &      2.1 &      0.5 &      0.8 \\ %   20  r 20964
   6 &   5.92 &   0.21 &   4.16 & $ -1.32$ & $ -1.19$ & $ -1.01$ &      2.6 &      0.7 &      0.8 \\ %   22  r 19866
   7 &   4.29 &   0.22 &   2.56 & $ -1.97$ & $ -1.84$ & $ -1.65$ &      8.8 &      1.7 &      2.7 \\ %   25  r 32810
   8 &   4.98 &   0.19 &   3.19 & $ -1.77$ & $ -1.63$ & $ -1.43$ &      9.8 &      3.0 &      0.9 \\ %   36  r 40728
   9 &   5.20 &   0.19 &   3.85 & $ -1.48$ & $ -1.35$ & $ -1.15$ &      5.4 &      2.0 &      1.3 \\ %   26  r 30630
  10 &   2.45 &   0.06 &   0.82 & $ -2.17$ & $ -2.14$ & $ -2.04$ &      6.3 &      1.6 &      1.9 \\ %   40  r 34557
  11 &   4.45 &   0.17 &   2.85 & $ -1.89$ & $ -1.75$ & $ -1.55$ &      6.4 &      1.7 &      2.0 \\ %   41  r 32361
  12 &   4.92 &   0.04 &   3.57 & $ -1.62$ & $ -1.48$ & $ -1.28$ &      6.1 &      2.1 &      2.7 \\ %    2 rw 200980
  13 &   5.43 &   0.08 &   3.81 & $ -1.50$ & $ -1.37$ & $ -1.17$ &      2.1 &      0.5 &      0.8 \\ %   27  r 27421
  14 &   5.81 &   0.38 &   3.98 & $ -1.42$ & $ -1.28$ & $ -1.10$ &      2.7 &      0.6 &      0.8 \\ %    3  r 20953
  15 &   4.20 &   0.08 &   2.84 & $ -1.89$ & $ -1.76$ & $ -1.55$ &      9.4 &      2.2 &      2.4 \\ %   15 rw 201004
  16 &   5.30 &   0.05 &   3.76 & $ -1.53$ & $ -1.39$ & $ -1.19$ &      5.2 &      1.1 &      2.1 \\ %   13 rw 201012
  17 &   4.04 &   0.03 &   2.60 & $ -1.96$ & $ -1.83$ & $ -1.63$ &      4.2 &      1.1 &      1.6 \\ %   42  r 30858
  18 &   6.00 &   0.23 &   4.16 & $ -1.32$ & $ -1.19$ & $ -1.01$ &      2.0 &      0.4 &      0.3 \\ %   24  r 34381
  19 &   5.25 &   0.02 &   3.72 & $ -1.55$ & $ -1.41$ & $ -1.21$ &      5.6 &      1.5 &      2.6 \\ %   39  r 36120
  20 &   4.95 &   0.20 &   3.26 & $ -1.74$ & $ -1.60$ & $ -1.40$ &      1.9 &      0.3 &      0.4 \\ %   34  r 27851
  21 &   5.47 &   0.51 &   3.51 & $ -1.64$ & $ -1.50$ & $ -1.30$ &      3.4 &      0.7 &      0.9 \\ %   17  r 26320
  22 &   5.02 &   0.14 &   3.59 & $ -1.61$ & $ -1.47$ & $ -1.27$ &     11.2 &      1.7 &      1.4 \\ %    8  r 12465
  23 &   6.20 &   0.03 &   4.34 & $ -1.22$ & $ -1.10$ & $ -0.92$ &      4.3 &      1.1 &      1.9 \\ %   21  r 19530
  24 &   5.70 &   0.13 &   4.17 & $ -1.32$ & $ -1.19$ & $ -1.00$ &      3.3 &      1.0 &      1.5 \\ %   38  r 37584
  25 &   5.11 &   0.39 &   3.67 & $ -1.57$ & $ -1.43$ & $ -1.23$ &      4.1 &      1.9 &      2.0 \\ %   10 rw 201168
  26 &   4.83 &   0.11 &   3.50 & $ -1.64$ & $ -1.51$ & $ -1.30$ &      5.8 &      2.0 &      2.1 \\ %   30 rw 201258
  27 &   5.57 &   0.02 &   3.83 & $ -1.49$ & $ -1.36$ & $ -1.16$ &      5.7 &      1.7 &      1.8 \\ %   33 rw 201269
  28 &   5.40 &   0.29 &   3.82 & $ -1.50$ & $ -1.36$ & $ -1.17$ &      4.6 &      1.3 &      1.5 \\ %    7  r 13814
  29 &   4.93 &   0.05 &   3.12 & $ -1.80$ & $ -1.66$ & $ -1.45$ &  \nodata &  \nodata &  \nodata \\ %    1  r 25974
  30 &   2.24 &   0.00 &   0.75 & $ -2.17$ & $ -2.14$ & $ -2.05$ &     10.2 &      2.0 &      1.0 \\ %   14 rw 201382
  31 &   5.36 &   0.46 &   3.72 & $ -1.55$ & $ -1.41$ & $ -1.21$ &      3.2 &      0.7 &      1.4 \\ %   58  r 31733
  32 &   4.53 &   0.41 &   3.05 & $ -1.82$ & $ -1.68$ & $ -1.48$ &      6.3 &      2.0 &      3.0 \\ %   46  r 23503
  33 &   4.90 &   0.01 &   3.10 & $ -1.80$ & $ -1.66$ & $ -1.46$ &      8.5 &      2.0 &      2.2 \\ %   31 rw 201416
  34 &   5.13 &   0.29 &   3.29 & $ -1.73$ & $ -1.59$ & $ -1.39$ &      1.7 &      0.2 &      0.3 \\ %    6  r 17954
  35 &   5.73 &   0.03 &   3.91 & $ -1.45$ & $ -1.32$ & $ -1.13$ &      4.9 &      1.2 &      1.7 \\ %    9  r 12705
  36 &   5.46 &   0.07 &   3.57 & $ -1.61$ & $ -1.48$ & $ -1.27$ &      2.6 &      0.5 &      0.8 \\ %   59  r 29925
  37 &   5.50 &   0.01 &   3.71 & $ -1.55$ & $ -1.41$ & $ -1.22$ &     10.2 &      2.5 &      1.9 \\ %   32 rw 201470
  38 &   5.40 &   0.33 &   3.68 & $ -1.57$ & $ -1.43$ & $ -1.23$ &      7.1 &      1.9 &      1.8 \\ %   16 rw 201479
  39 &   5.32 &   0.48 &   3.82 & $ -1.50$ & $ -1.36$ & $ -1.17$ &      2.5 &      0.7 &      1.3 \\ %    5  r 21033
  40 &   5.15 &   0.11 &   3.66 & $ -1.57$ & $ -1.44$ & $ -1.24$ &      7.2 &      2.0 &      2.7 \\ %   11 rw 201535
  41 &   5.08 &   0.05 &   3.21 & $ -1.76$ & $ -1.62$ & $ -1.42$ &      5.5 &      1.3 &      2.0 \\ %   49  r 18263
  42 &   6.45 &   0.03 &   4.34 & $ -1.22$ & $ -1.10$ & $ -0.92$ &  \nodata &  \nodata &  \nodata \\ %   57  r 33719
  43 &   6.12 &   0.12 &   4.45 & $ -1.16$ & $ -1.04$ & $ -0.87$ &      2.8 &      0.8 &      1.2 \\ %   44  r 13082
  44 &   5.28 &   0.21 &   3.44 & $ -1.67$ & $ -1.53$ & $ -1.33$ &      2.6 &      0.7 &      0.7 \\ %   54  r 22971
  45 &   2.94 &   0.23 &   1.52 & $ -2.16$ & $ -2.07$ & $ -1.92$ &      4.9 &      1.5 &      1.9 \\ %   56  r 38168
  46 &   6.13 &   0.07 &   4.33 & $ -1.23$ & $ -1.10$ & $ -0.93$ &      3.8 &      1.1 &      1.3 \\ %   50  r 17061
  47 &   5.86 &   0.44 &   4.07 & $ -1.37$ & $ -1.24$ & $ -1.05$ &      1.5 &      0.1 &      0.2 \\ %   53  r 24744
  48 &   6.21 &   0.21 &   4.51 & $ -1.12$ & $ -1.00$ & $ -0.84$ &      1.7 &      0.2 &      0.2 \\ %   52  r 27889
  49 &   3.26 &   0.26 &   1.41 & $ -2.16$ & $ -2.09$ & $ -1.95$ &      8.0 &      1.9 &      1.9 \\ %   43  r 26633
  50 &   5.81 &   0.04 &   3.65 & $ -1.58$ & $ -1.44$ & $ -1.24$ &  \nodata &  \nodata &  \nodata \\ %   51  r 15860
  51 &   4.96 &   0.06 &   3.09 & $ -1.81$ & $ -1.67$ & $ -1.47$ &      7.8 &      1.3 &      1.0 \\ %   55  r 41546
  52 &   5.79 &   0.38 &   3.89 & $ -1.46$ & $ -1.33$ & $ -1.13$ &      4.0 &      1.0 &      1.1 \\ %   45  r 24711
  53 &   5.67 &   0.11 &   3.96 & $ -1.43$ & $ -1.29$ & $ -1.10$ &      4.6 &      1.4 &      2.4 \\ %   48  r 19509
  54 &   6.17 &   0.22 &   4.55 & $ -1.09$ & $ -0.98$ & $ -0.81$ &      1.9 &      0.3 &      0.7 \\ %   47  r 20668
 } 
{Measurements of metallicity and age for stars in \leoi. 
}{%put the label
  t:data}{%size
  normalsize}{
  {\it Notes}.  
  $\epsilon\,_{\Sigma W}$ is the absolute semi-difference of the
  equivalent widths measured on the individual spectra.  The listed
  metallicity values were calculated from $W^\prime$ using the
  quadratic calibration. The last two columns give the lower and upper
  confidence intervals of our age estimates (see text for details). }

%++++++++++++++++++++++++++++++++++++++++++++++++++++++++++++++++++++++++
\end{document}